\documentclass[12pt,draftclsnofoot, perreview, onecolumn]{IEEEtran}%Prof. Ma's template
\usepackage{bbm}
\usepackage{color}
\usepackage{amsmath}
\usepackage{mathrsfs}
\usepackage{pifont}
\usepackage{amsfonts}
\usepackage{citesort}
\usepackage{graphicx}
\ifCLASSINFOpdf
\else
\fi

\newtheorem{theorem}{Theorem}
\newtheorem{definition}{Definition}
\newtheorem{algorithm}{Algorithm}
\newtheorem{example}{Example}
\newtheorem{corollary}{Corollary}
\newtheorem{lemma}{Lemma}

\begin{document}
\title{Accessible Capacity of Secondary Users}

\author{Authors}

\author{Xiujie~Huang,~\IEEEmembership{Member,~IEEE,}
        Xiao~Ma,~\IEEEmembership{Member,~IEEE,}
        Lei~Lin, and~Baoming~Bai,~\IEEEmembership{Member,~IEEE}
\thanks{This work was partially presented at ISIT2011.

Manuscript received....This work was supported by the National Basic Research Program
of China (973 Program, No.~2012CB316100) and by the National Science Foundation of China (No.~61172082).}
\thanks{X.~Huang was with the Department of Electronics and Communication Engineering, Sun Yat-sen University, Guangzhou 510006, Guangdong, China. She is currently with the Department of Electrical Engineering, University of Hawaii, Honolulu 96822, HI, USA~(email: xiujie@hawaii.edu). }
\thanks{X.~Ma is with the Department of Electronics and Communication Engineering, Sun Yat-sen University, Guangzhou 510006, Guangdong, China~(email: maxiao@mail.sysu.edu.cn).}
\thanks{L.~Lin is with the Department of Mathematics, Sun Yat-sen University, Guangzhou 510275, Guangdong, China.}
\thanks{B.~Bai is with the State Lab. of ISN, Xidian University, Xi'an 710071, Shaanxi, China.}}

\markboth{SUBMITTED TO IEEE TRANSACTIONS ON INFORMATION THEORY,
December, 2010}{Huang \MakeLowercase{\textit{et~al.}}:Accessible Capacity of Secondary Users}




\maketitle\thispagestyle{empty}

%\begin{abstract}
%A new problem formulation is presented for the Gaussian interference channels~(GIFC) with two pairs of users, which are distinguished as primary users and secondary users, respectively. The primary users employ a pair of encoder and decoder that were originally designed to satisfy a given error performance requirement under the assumption that no interference exists from other users. In the case when the secondary users attempt to access the same medium, we are interested in the maximum transmission rate~(defined as {\em accessible capacity}) at which secondary users can communicate reliably without affecting the error performance requirement by the primary users under the constraint that the primary encoder (not the decoder) is kept unchanged. By modeling the primary encoder as a generalized trellis code~(GTC), we are then able to treat the secondary link as a finite state channel~(FSC). Upper and lower bounds on the accessible capacity are derived. For some special cases, these bounds can be computed numerically by using the BCJR algorithm. The numerical results show us, as expected, that primary users with lower transmission rates may allow higher accessible rates, and that better primary encoders guarantee not only higher quality of the primary link but also higher accessible rates of the secondary users. More interestingly,  the numerical results show that the accessible capacity does not always increase with the transmission power of the secondary transmitter.
%\end{abstract}

\begin{abstract}
A new problem formulation is presented for the Gaussian interference channels~(GIFC) with two pairs of users, which are distinguished as primary users and secondary users, respectively. The primary users employ a pair of encoder and decoder that were originally designed to satisfy a given error performance requirement under the assumption that no interference exists from other users. In the scenario when the secondary users attempt to access the same medium, we are interested in the maximum transmission rate~(defined as {\em accessible capacity}) at which secondary users can communicate reliably without affecting the error performance requirement by the primary users under the constraint that the primary encoder (not the decoder) is kept unchanged. By modeling the primary encoder as a generalized trellis code~(GTC), we are then able to treat the secondary link and the cross link from the secondary transmitter to the primary receiver as finite state channels~(FSCs). Based on this, upper and lower bounds on the accessible capacity are derived. The impact of the error performance requirement by the primary users on the accessible capacity is analyzed by using the concept of interference margin. In the case of non-trivial interference margin, the secondary message is split into common and private parts and then encoded by superposition coding, which delivers a lower bound on the accessible capacity. For some special cases, these bounds can be computed numerically by using the BCJR algorithm. Numerical results are also provided to gain insight into the impacts of the GTC and the error performance requirement on the accessible capacity.

%Two techniques are employed to derive upper and lower bounds for the accessible capacity. First, we treat both the secondary link and the cross link from the secondary transmitter to the primary receiver as finite state channels~(FSC) by modeling the primary encoder as a generalized trellis code~(GTC). Second, we use Han and Kobayashi's idea to split the secondary message into common and private parts.
%\textcolor{blue}{ As expected, the numerical results show us: 1)~primary users with lower transmission rates may allow higher accessible rates; 2)~better primary encoders guarantee not only higher quality of the primary link but also higher accessible rates for the secondary users; 3)~a greater error performance requirement allow a higher accessible rate. More interestingly, the numerical results show that the accessible rate does not always increase with the transmission power of the secondary transmitter.}
\end{abstract}


\begin{IEEEkeywords}
Accessible capacity, accessible rate, cognitive Gaussian interference channel, finite state channel~(FSC), Gaussian interference channel~(GIFC), generalized trellis code~(GTC), interference margin, limit superior in probability
\end{IEEEkeywords}

% \fi
\IEEEpeerreviewmaketitle

\section{Introduction}\label{SecIntr}

\subsection{Gaussian Interference Channel}

\IEEEPARstart{A}{s} an important model for wireless network communications, the Gaussian interference channel~(GIFC) was first mentioned by Shannon~\cite{Shannon61} in 1961 and studied a decade later by Ahlswede~\cite{Ahlswede74} who gave simple but fundamental inner and outer bounds on the capacity region of the GIFC. In 1978, Carleial~\cite{Carleial78} proved that any GIFC with two pairs of users can be standardized by scaling as
\begin{equation}\label{eqnGIFC}
            \begin{array}{ccrcrcc}
             Y_1 & = & X_1 & + & a_{21} X_2 & + & Z_1 \\
             Y_2 & = & a_{12} X_1 & + & X_2 & + & Z_2
           \end{array},
\end{equation}
where the real numbers $X_i \in \mathcal{X}_i$, $Y_i \in \mathcal{Y}_i$ and $Z_i \in \mathbb{R}$~($i\in\{1,2\}$) are the channel inputs, outputs and additive noises, respectively. The channel inputs $X_i$ are required to satisfy power constraints $P_i$ and the noises $Z_i$ are samples from a white Gaussian process with double-sided power spectrum density one. The GIFC is then completely specified by the interference coefficients $a_{12}$ and $a_{21}$ as well as the transmission powers $P_1$ and $P_2$. Carleial also showed that, in the case that the interference is very strong~(i.e., $a_{12}^2 \geq 1 + P_2$ and $a_{21}^2 \geq 1 + P_1$), the capacity region is a rectangle~\cite{Carleial75,Carleial78}. When the interference is strong~(i.e., $a_{12}^2 \geq 1$ and $a_{21}^2 \geq 1$), Han and Kobayashi~\cite{Han81}, and Sato~\cite{Sato78} obtained the capacity region by transforming the original problem into the problem to find the capacity region of a compound multiple-access channel. The idea of this transformation was also employed to find the capacity regions of another class of GIFCs, where the channel outputs $Y_1$ and $Y_2$ are statistically equivalent~\cite{Ahlswede74,Carleial78}.

However, the determination of the capacity region of the general GIFC is still open. Only various inner and outer bounds are presented. Among these, the best inner bound is that put forth by Han and Kobayashi~\cite{Han81}, which has been simplified by Chong {\em et~al.} and Kramer in their independent works~\cite{Chong08} and~\cite{Kramer06}. In 2004, Kramer derived two outer bounds on the capacity region of the general GIFC~\cite{Kramer04}. The first bound for general GIFC unifies and improves the outer bounds of Sato~\cite{Sato77} and Carleial~\cite{Carleial83}. The second bound, which is based on one-sided GIFCs~(i.e., $a_{12} = 0$ or $a_{21}=0$), follows directly from the outer bounds of Sato~\cite{Sato78} and Costa~\cite{Costa85}, and is proved to be better than the first one for certain weak GIFCs~(i.e., $0 < a_{12}^2 \leq 1$ or $0 < a_{21}^2 \leq 1$).

In light of the difficulty in finding the exact capacity regions of general GIFCs, Etkin {\em et~al.}~\cite{Etkin08} introduced the idea of approximation to show that Han and Kobayashi's inner bound~\cite{Han81} is within one bit of the capacity region, and also gave outer bounds for the weak GIFC~(i.e., $0 < a_{12}^2 \leq 1$ and $0 < a_{21}^2 \leq 1$) and the mixed GIFC~(i.e., $0 < a_{12}^2 \leq 1$ and $a_{21}^2 \geq 1$, or $a_{12}^2 \geq 1$ and $0 < a_{21}^2 \leq 1$). This fresh approximation approach is recently widely used in understanding and exploring multiuser Gaussian channels. For example, Bresler {\em et~al.}~\cite{Bresler10} extended the approximation method to investigate the capacity regions of many-to-one and one-to-many GIFCs and showed that the capacity regions can be determined to within constant gaps. They also proposed the use of lattice codes for alignment of interfering signals on the signal level instead of in the signal space~\cite{Cadambe08,Maddah-Ali08}. However, for the two user GIFC, the result of being within one bit of the capacity region is particularly relevant in the high signal-to-noise ratio~(SNR) regime. In~\cite{Motahari09}, Motahari and Khandani proposed upper bounds for weak GIFC and mixed GIFC, both of which outperform the upper bounds of Kramer~\cite{Kramer04} and Etkin {\em et~al.}~\cite{Etkin08}.

\subsection{Cognitive Radio}

GIFC is also an important model to explore the throughput potential of cognitive radios. In the scenarios of cognitive radios~\cite{Haykin05}, the secondary~(unlicensed) users are allowed to communicate with each other using the same spectrum as allocated to the primary~(licensed) users provided that their communications do not interfere the primary users. As pointed out by Srinivasa and Jafar~\cite{Srinivasa06}, cognitive radios can be classified in a very broad sense into three groups, which seek to underlay, interweave or overlay the secondary users' signals with the primary users' signals in such a way that the primary users of the spectrum are as unaffected as possible.

In underlay cognitive radios, the secondary users are assumed to be capable of measuring the current radio environment, and adjusting their transmission characteristics~(say, spreading their transmission power in an ultra-wide band) in such a way that the interference temperature at the primary receivers remains below a preset limit. With this constraint, Clancy~\cite{Clancy07} has developed a model to analyze interference temperature and to examine the relationships between the capacity achieved by the secondary users and the interference caused to the primary users. In interweave cognitive radios, the secondary users periodically monitor the radio spectrum, detect the presence/absence of the primary users and then opportunistically interweave their transmitted signals through gaps that arise in frequency and time. For such cognitive radios, key problems include resource allocation, cognition and sensing. Capacity limits with various models for resource allocation, cognition and sensing were investigated in~\cite{Jafar07,Chung10} and the references therein. In overlay cognitive radios, the secondary users transmit signals simultaneously with the primary users. The overlay cognitive radio channel~(also referred to as cognitive GIFC) is very similar to the classical GIFC. The main difference lies in that the secondary users are assumed to know~(non-causally or causally) full (or partial) messages to be transmitted at the primary users~\cite{Devroye06}. This assumption is asymmetric, hence the overlay cognitive radio channels~\cite{Chung12} are also known as GIFCs with unidirectional cooperation~\cite{Maric06,Maric07} or GIFCs with degraded message sets~\cite{Wu07}. Like classical GIFCs, the capacity region of the cognitive GIFC is only known in certain parameter regimes but remains unknown in general, say, capacity regions were determined in~\cite{Maric07} for a special channel with ``strong interference'' under certain assumptions and in~\cite{Wu07,Jovicic09} for a class of channels with ``weak interference" at the primary receiver. Instead, inner and outer bounds with various assumptions have been proposed, for more details see~\cite{Rini11,Rini12} and the references therein.

All the above techniques rely mainly on the transmitter-side cognition and/or cooperation. Popovski~{\it et~al.}~\cite{Popovski07} focused on receiver-side cognition and proposed opportunistic interference cancellation~(OIC) at the secondary receiver. The idea is as follows. The secondary receiver monitors the data rate as well as the received power from the primary transmitter and checks if the primary signal~(interference) is decodable. Once such an opportunity occurs, the secondary receiver can inform the secondary transmitter to adjust the transmission rate such that the noisy multiple access signals at the secondary receiver can be decoded by first decoding and canceling the primary signals. Popovski~{\em et~al.}~\cite{Popovski07} also devised a method~(superposition coding) to achieve the maximum achievable rate of the secondary user in this setting. The dependence of outage probabilities on the channel state information with OIC~(or suboptimal OIC) has been analyzed in~\cite{DiTaranto11}. The idea of receiver-side opportunism, which has been extended to cognitive networks with multiple-secondary-users~\cite{Devroye11}, has a practical significance since it requires less cooperations from the primary~(legitimate) users.

%-------------------------------------------------------------------------------%
\begin{figure}
  \centering
  \includegraphics[width=3.5in]{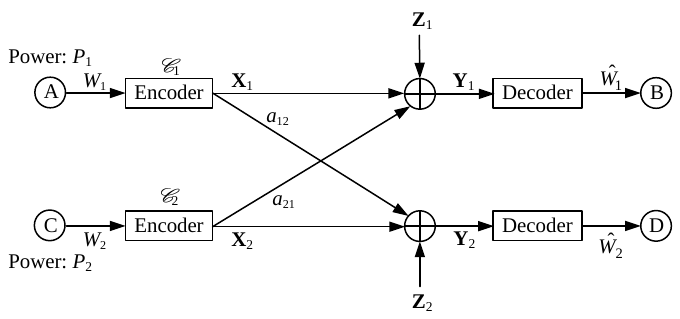}\\
  \caption{The system model of a Gaussian interference channel.}\label{FigGIFCC} % with primary and secondary users
\end{figure}
%-------------------------------------------------------------------------------%

\subsection{New Problem Formulation}

In this paper, we present a new problem formulation for the GIFC with two users. For ease of comparisons, let us recall the original information theoretic problem of the GIFC. The whole system with codes $\mathscr{C}_1$ and $\mathscr{C}_2$ is shown in Fig.~\ref{FigGIFCC}. Briefly, the system works as follows. The messages at User A are encoded by $\mathscr{C}_1$ and transmitted to User B, while the messages at User C are encoded by $\mathscr{C}_2$ and transmitted to User D. The messages from User A and C are assumed to be independent and the two senders do not collaborate with each other. User B and D work independently to decode the respective received signals for the purpose of correctly extracting the respective messages. User B and D are assumed to know exactly the structures of $\mathscr{C}_1$ and $\mathscr{C}_2$. The problem of finding the capacity region is equivalent to that of determining whether or not a pair of codes $(\mathscr{C}_1, \mathscr{C}_2)$ exist with any given respective rates $(R_1, R_2)$ such that the decoding error probabilities are arbitrarily small. In this original formulation, both $\mathscr{C}_1$ and $\mathscr{C}_2$ are allowed to be varied to determine the limits. Typically, on~(or near) the boundary of the capacity region, they must be an optimal~(or near-optimal) pair of codes.

In our new formulation, we follow the terminology in cognitive GIFCs and distinguish the two pairs of users as primary users and secondary users, respectively. The primary users~(User A and B) employ a pair of encoder $\mathscr{C}_1$ and decoder that were originally designed to satisfy a given error performance requirement under the assumption that no interference exists from other users. In the case when the secondary users~(User C and D) attempt to access the same medium, we are interested in the maximum transmission rate~(defined as {\em accessible capacity}) at which secondary users can communicate reliably without affecting the error performance requirement by the primary users under the constraint that the primary encoder (not the decoder) is kept unchanged. That is, we make an assumption that the code $\mathscr{C}_1$ of rate $R_1$ is fixed and only the code $\mathscr{C}_2$ is allowed to be varied for the purpose of maximizing the coding rate $R_2$. This assumption is reasonable at least in the following three scenarios.
\begin{itemize}
    \item It is not convenient~(or economic) to change the encoder $\mathscr{C}_1$  at User A for dealing with the interference from the secondary users. For example, User A is located in a place~(say the Space Station) that cannot be reached easily.
    \item User A is weak in the sense that it can only afford the simple encoders such as $\mathscr{C}_1$ due to the limit of its processing ability. For example, User A is an energy-limited wireless sensor that collects and transmits data to the powerful data center~(User B).
    \item This assumption becomes more reasonable in the cognitive radios since the primary encoder is a part of a legacy system and the primary user, as a legitimate user, may not want to change the encoder.
\end{itemize}

Apparently, due to the constraint that the primary encoder was designed for interference-free channels and cannot be changed, our formulation is different from both the classical GIFCs and the cognitive GIFCs. The detailed difference will be given in Sec.~\ref{SecProb} after the model is explicitly defined.

\subsection{Structure and Notations}

The main results as well as the structure of this paper are summarized as follows.
\begin{enumerate}
  %\item In Sec.~\ref{SecIntr}, as well as related results of classical GIFCs and cognitive GIFCs, the motivation of our new problem formulation is introduced.
  \item In Sec.~\ref{SecProb}, the accessible capacity is explicitly defined by modeling the primary encoder as a {\em generalized trellis code}~(GTC) and using the concept of {\em limit superior in probability} introduced in~\cite{Han93}. The relationships between our new formulation and existing works are also revealed in this section.
  \item In Sec.~\ref{SecBoundCap}, upper and lower bounds on accessible capacities are derived by treating the secondary link and the cross link from the secondary transmitter to the primary receiver as finite state channels~(FSCs)~\cite{Gallager68}. To investigate the impact of the error performance requirement by the primary users on the accessible capacity, we borrow the idea of interference margin from~\cite{Devroye11}. When non-trivial interference margin exists, we propose a lower bound by using superposition coding.
  \item In Sec.~\ref{SubSecComRats}, we show that the derived bounds on the accessible capacity can be evaluated numerically using the BCJR algorithm~\cite{BCJR74} for special cases.
  \item In Sec.~\ref{SubSecNumRes}, simulations are presented, showing numerically the following either expected or interesting results.
      \begin{itemize}
        \item Primary users with lower transmission rates may allow higher accessible rates.
        \item Better primary encoders guarantee not only higher quality of the primary links but also higher accessible rates of the secondary users.
        \item The accessible rate does not always increase with the transmission power of the secondary transmitter.
        \item Relaxing the quality requirement by the primary users allows higher accessible rates for the secondary users.
      \end{itemize}
  \item In Sec.~\ref{SecConclusion}, we conclude our work.
\end{enumerate}

In this paper, a random variable is denoted by a capital letter, say $X$, while its realization and sample space are denoted by the lower-case letter $x$ and $\mathcal{X}$, respectively. A sequence of random variables $(X_1, X_2, \cdots, X_N)$ is denoted by ${\bf X}$. The probability mass function~(pmf) of a discrete random variable $X$ is denoted by $p_X(x)$, while the probability density function~(pdf) of a continuous random variable $Y$ is denoted by $f_Y(y)$. The transition probability mass~(or density) function from $X$ to $Y$ is denoted by $p_{Y|X}(y|x)$~(or $f_{Y|X}(y|x)$). To avoid cluttering the notation in some contexts, we may use, for example, $p(x_1)$ in place of $p_{X_1}(x_1)$ and $f(y_1|x_2)$ in place of $f_{Y_1|X_2}(y_1|x_2)$.

\section{Basic Definitions and Problem Statements}\label{SecProb}

\subsection{Interference-Free AWGN Channels}\label{SubSecPL}

Referring to Fig.~\ref{FigGIFCC}, we assume that only primary users, User A and B, exist at the beginning. That is, User A is sending  messages to User B through a discrete-time AWGN channel without any interference from other users. The messages from User A are usually represented by integers and required to be coded and modulated as a sequence of real signals. This  process can be described in a unified way by introducing the concept of {\em generalized trellis code~(GTC)} as follows.
\begin{itemize}
  \item The code can be represented by a {\em time-invariant} trellis and (hence) is uniquely specified by a trellis section.
  \item A trellis section is composed of {\em left states} and {\em right states} which are connected by {\em branches} in between. Both the left and right states are selected from the same set $\mathcal{S} = \{0, 1, \ldots, |\mathcal{S}| -1\}$.
  \item Emitting from each state there are $M$ branches. A branch is specified by a four-tuple $b \stackrel{\Delta}{=} (s^{-}(b), u(b), c(b), s^{+}(b))$, where $s^{-}(b)$ is the starting state, $s^{+}(b)$ is the ending state, $u(b) \in \{0, 1, \ldots, M-1\}$ is an integer that represents a message to be encoded, and $c(b) \in \mathbb{R}^n$ is an $n$-dimensional real signal to be transmitted over the channel. We assume that a branch $b$ is uniquely determined by $s^{-}(b)$ and $u(b)$. We denote the collection of all branches by $\mathcal{B}$. The coding rate is $R_{GTC} \stackrel{\Delta}{=} \frac{\log M}{n}$.
  \item Without loss of generality, we assume that the average energy emitted from each state is normalized, i.e., $\frac{1}{M} \sum_{b: s^{-}(b) = s} \|c(b)\|^2 = n$ for all $s$, where $\|c(b)\|$ represents the squared Euclidean norm of $c(b)$.
\end{itemize}

%------------------------------------------------------------------------------%
\begin{figure}
  \centering
  \includegraphics[width=13.0cm]{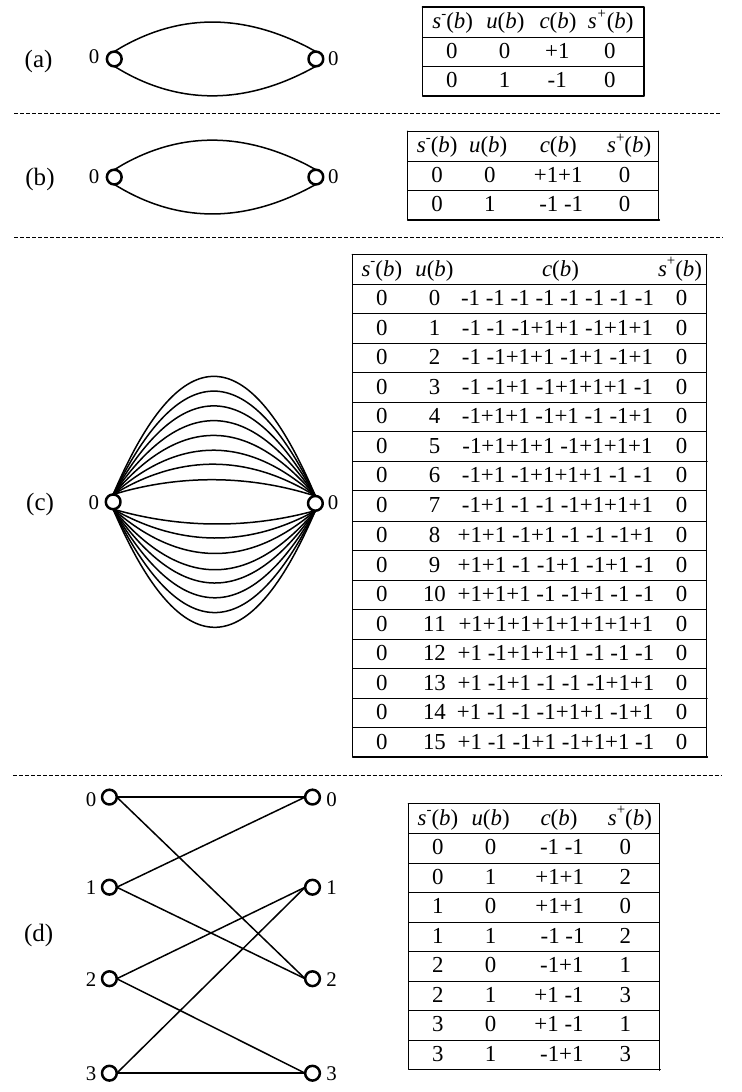}\\
  \caption{Generalized trellis codes. (a)~Uncoded BPSK. (b)~Repetition coded BPSK. (c)~Extended Hamming coded BPSK.
  (d)~Convolutional coded BPSK.}\label{FigGTCs}
\end{figure}
%------------------------------------------------------------------------------%

To help readers understand the concept of GTC, we give four examples below.

\begin{example}[Uncoded BPSK]\label{ExBPSK}
The binary phase shift keying~(BPSK) modulation can be considered as a GTC. The trellis section is composed of one left state and one right state which are connected by two parallel branches. The two branches encode messages $0$ and $1$ to $-1$ and $+1$, respectively. The trellis section and the branch set $\mathcal{B}$ are shown in Fig.~\ref{FigGTCs}~(a).
\hfill \ding{113}
\end{example}

\begin{example}[Repetition Coded BPSK~(RCBPSK)]\label{ExRCBPSK}
The simplest repetition code $[2,1,2]$ of rate $1/2$ with the BPSK signaling can be regarded as a GTC. The trellis section is composed of one left state and one right state which are connected by two parallel branches. The two branches encode messages $0$ and $1$ to $(-1, -1)$ and $(+1, +1)$, respectively. Fig.~\ref{FigGTCs}~(b) gives the trellis representation of this GTC.
\hfill \ding{113}
\end{example}

\begin{example}[Extended Hamming Coded BPSK~(EHCBPSK)]\label{ExEHCBPSK}
Consider the $[8, 4, 4]$ extended Hamming code defined by the parity-check matrix
\begin{equation}
    {\bf H} = \left(\begin{array}{cccccccc}
                      1 & 0 & 1 & 1 & 1 & 0 & 0 & 0 \\
                      0 & 1 & 0 & 1 & 1 & 1 & 0 & 0 \\
                      0 & 0 & 1 & 0 & 1 & 1 & 1 & 0 \\
                      1 & 1 & 1 & 1 & 1 & 1 & 1 & 1
                    \end{array}
    \right).
\end{equation}
The extended Hamming code with the BPSK signaling can be regarded as a GTC. The trellis section is composed of one left state and one right state which are connected by sixteen parallel branches, each of which encodes an integer~(of four binary digits) to an eight-dimensional real signal. Fig.~\ref{FigGTCs}~(c) depicts the trellis representation of this GTC.
\hfill \ding{113}
\end{example}

\begin{example}[Convolutional Coded BPSK~(CCBPSK)]\label{ExCCBPSK}
Consider the $(2,1,2)$ convolutional code defined by the generator matrix
\begin{equation}
    G(D) = [1 + D^2 \;\;\; 1+D+D^2].
\end{equation}
The convolutional code with the BPSK signaling can be regarded as a GTC. At each stage of the trellis, there are four states $\{0,1,2,3\}$, and from each state there are two branches, each of which encodes a binary digit to a two-dimensional real signal. The trellis section and the branch set $\mathcal{B}$ are shown in Fig.~\ref{FigGTCs}~(d).
\hfill \ding{113}
\end{example}

The system model for the primary link with a GTC is described as follows.

{\em \underline{Encoding}:} Let $w_1 = (u_1, u_2, \ldots, u_N)\in \{0, 1, \ldots, M-1\}^N$ be an $M$-ary data sequence, drawn from an independent and uniformly distributed~(i.u.d.) source, to be transmitted. Denote $\mathcal{M}_1 \stackrel{\Delta}{=} \{0, 1, \ldots, M-1\}^N$ and call it ``super'' message set. Obviously, $|\mathcal{M}_1|= M^N$. If necessary, $N$ is allowed to be sufficiently large. The encoding is described as follows.
\begin{enumerate}
    \item[1)] At time $t =0$, the state of the encoder is initialized as $s_0 \in \mathcal{S}$.
    \item[2)] At time $t = 1, 2, \ldots$, the message $u_t$ is input to the encoder and drives the encoder from state $s_{t-1}$ to $s_t$. In the meantime, the encoder delivers a coded signal $c_t$ such that $(s_{t-1}, u_t, c_t, s_t)$ forms a valid branch.
    \item[3)] Suppose that the available power is $P_1$. Then the signal $x_{1, t} = \sqrt{P_1} c_t$ at time $t$ is transmitted. The transmitted signal sequence is denoted by ${\bf x}_1$. The collection of all coded~(transmitted) sequences is denoted by $\mathscr{C}_1$. Notice that $\mathscr{C}_1$ may depend on $s_0$. We assume that, given $s_0$, all transmitted sequences are distinct.
\end{enumerate}

{\em \underline{AWGN Channel}:} The channel is assumed to be an AWGN channel and the received signal sequence is denoted by ${\bf y}_1$, which is statistically determined by
 \begin{equation}\label{AWGN}
    {\bf y}_1 = {\bf x}_1 + {\bf z}_1
 \end{equation}
where ${\bf z}_1$ is a sequence of samples from a white Gaussian noise of variance one per dimension.

{\em \underline{Decoding}:} Upon receiving ${\bf y}_1$, User B can utilize, in principle, the Viterbi algorithm~\cite{Forney72}, the BCJR algorithm~\cite{BCJR74} or other trellis decoding algorithms~\cite{Ma03} to estimate the transmitted messages. Assume that a decoder $\psi_1$ is utilized and $\hat{w}_1 = (\hat{u}_1, \hat{u}_2, \ldots, \hat{u}_N)$ is the estimated message sequence after decoding.

{\em \underline{Error Performance Criterion}:}
For all $t \geq 1$, define error random variables as
\begin{equation}\label{eqnErrorseq}
    E_t = \left\{\begin{array}{cc}
                   0, & {\rm~if~} \hat{U}_t = U_t \\
                   1, & {\rm~if~} \hat{U}_t \neq U_t
                 \end{array}
            \right..
\end{equation}
Depending on the structure of the GTC as well as the assumed decoding algorithm $\psi_1$, the statistical dependence among the random variables $\{E_t\}$ may be very complicated.
In order to characterize the performance of the (de)coding scheme in a unified way, we introduce the following random variables
\begin{equation}\label{eqnDefThetaN}
    \Theta_N = \frac{\Sigma_{t=1}^N E_t}{N}, {\rm~for~} N = 1,2,\ldots
\end{equation}
and consider the {\em limit superior in probability}~\cite{Han93} of the sequence $\{\Theta_N\}$.

\begin{definition}\label{DefSatisfactory}
Let $\varepsilon$ be a real number in the interval $(0,1)$. A GTC is said to be $\varepsilon$-{\em satisfactory} under the decoder $\psi_1$ if the limit superior in probability of $\{\Theta_N\}$ is not greater than $\varepsilon$, that is,
\begin{equation}\label{pLimSupErr}
    p\texttt{-}\limsup \Theta_N \stackrel{\Delta}{=} \inf \left\{\alpha \, | \lim_{N\rightarrow \infty} {\rm Pr} \{\Theta_N > \alpha \} = 0 \right\} \leq \varepsilon.
\end{equation}
%Equivalently,
%\begin{equation}
%    \lim_{N\rightarrow \infty} {\rm Pr} \{\Theta_N > \varepsilon \} = 0.
%\end{equation}
\hfill \ding{113}
\end{definition}

In this paper, the given real number $\varepsilon$ is referred to as the {\em error performance requirement} by the primary users.

{\bf Remarks.}
\begin{enumerate}
  \item From Examples~\ref{ExRCBPSK} and~\ref{ExEHCBPSK}, we can see that a conventional block code of size $M$ can be regarded as a GTC. The trellis section has only one state and $M$ parallel branches, which correspond to $M$ codewords, respectively. Such a representation is different from those conventional trellis representations in~\cite{BCJR74,McEliece96,Vardy98}. For this special class of GTCs, the error random variables $\{E_t\}$ under commonly-used decoders are independent and identically distributed~(i.i.d.). Then, by the weak law of large numbers, we know that the word-error-rate~(WER) $\Theta_N$ converges to the expectation of $E_1$ in probability. That is, for any $\delta > 0$,
    \begin{equation}
        \lim_{N \rightarrow \infty} {\rm Pr} \left\{\left| \Theta_N - \lambda_1 \right| \leq \delta \right\} = 1
    \end{equation}
    where $\lambda_1 = {\rm Pr}(E_1 = 1)$ is the word error probability~(WEP) of the block code. In this case, $p\texttt{-}\limsup \Theta_N = \lambda_1$ and the definition that a block code is said to be $\varepsilon$-satisfactory is equivalent to saying that the WEP is not higher than $\varepsilon$.

  \item For the convolutional code given in Example~\ref{ExCCBPSK}, the error random variables $\{E_t\}$ under commonly-used decoders are usually dependent. In this case, $p\texttt{-}\limsup \Theta_N$ characterizes the limiting behavior of the bit-error-rate~(BER) $\Theta_N$.
\end{enumerate}

\subsection{Gaussian Interference Channels}\label{SubSecPLSL}

Referring to Fig.~\ref{FigGIFCC} again, we assume that User C attempts to send messages to User D by accessing the same medium as used by the primary users. In this scenario, cross-talks~(interferences) may occur. Assume that it is not convenient to change the encoder at User A. Now an interesting question arises: What is the maximum~(reliable) transmission rate from User C to User D under the constraint that the encoder at User A remains unchanged but the error performance requirement is still fulfilled? The detailed formulation of this problem is presented in the following.

{\em \underline{Encoding}:} The message set at User~A, as mentioned in the previous subsection, is set to be the ``super'' set $\mathcal{M}_1=\{0, 1, \ldots, M-1\}^N$, while the message set at User~C is set to be $\mathcal{M}_2 = \{1,2, \cdots, M_2\}$.
\begin{enumerate}
    \item[1)] The encoding function at User A is
        \begin{eqnarray}
            \begin{array}{cccl}
                \phi_1: & \mathcal{M}_1 & \rightarrow & \mathbb{R}^{nN} \\
                   & w_1  &\mapsto & {\bf x}_1 = \phi_1(w_1)
            \end{array},
        \end{eqnarray}
        where message $w_1 = (u_1, u_2, \ldots, u_N)\in \mathcal{M}_1$ is an $M$-ary sequence drawn from an i.u.d.~source and ${\bf x}_1$ is the coded sequence of length $nN$ such that $(w_1, {\bf x}_1)$ corresponds to a path through the trellis of the GTC $\mathscr{C}_1$.
        As in the absence of secondary users, the coded sequence ${\bf x}_1$ satisfies $\mathbf{E}\left[\|{\bf X}_1\|^2\right] = nNP_1$.

    \item[2)] The encoding function at User C is
        \begin{eqnarray}
            \begin{array}{cccl}
                \phi_2: & \mathcal{M}_2 & \rightarrow & \mathbb{R}^{nN} \\
                   & w_2  &\mapsto & {\bf x}_2 = \phi_2(w_2)
            \end{array},
        \end{eqnarray}
        where message $w_2$ is an integer uniformly distributed over $\mathcal{M}_2$, and ${\bf x}_2$ is the coded sequence of length $nN$ under the power constraint $\mathbf{E}\left[\|{\bf X}_2\|^2\right] \leq nNP_2$.

    \item[3)] The coding rates~(bits/dimension) at User A and C are $R_1 = R_{GTC} = \frac{\log M}{n}$ and $R_2 \stackrel{\Delta}{=} \frac{\log M_2}{nN}$, respectively.
\end{enumerate}

{\em \underline{Gaussian Interference Channels}:} Assume that User A and C transmit synchronously ${\bf x}_1$
and ${\bf x}_2$, respectively. The received sequences at User B and D are ${\bf y}_1$ and ${\bf y}_2$, respectively.
For the standard GIFC shown in Fig.~\ref{FigGIFCC}, we have
\begin{equation}\label{eqnIFCseq}
           \begin{array}{ccrcrcl}
            {\bf y}_1 & = & {\bf x}_1 & + & a_{21} {\bf x}_2 & + & {\bf z}_1\\
            {\bf y}_2 & = & a_{12} {\bf x}_1 & + & {\bf x}_2 & + & {\bf z}_2
           \end{array},
\end{equation}
where ${\bf z}_1$ and ${\bf z}_2$ are two sequences of samples drawn from an AWGN of variance one per dimension, and ${\bf a} = (a_{12}, a_{21})$ is the real interference coefficient vector.

{\em \underline{Decoding}:}
\begin{enumerate}
    \item[1)] The decoding function at User B is
        \begin{eqnarray}
            \begin{array}{cccl}
                {\tilde \psi}_1: & \mathbb{R}^{nN} & \rightarrow & \mathcal{M}_1 \\
                   & {\bf y}_1  &\mapsto & \tilde{w}_1 \stackrel{\Delta}{=} (\tilde{u}_1, \tilde{u}_2, \ldots, \tilde{u}_N) = {\tilde \psi}_1({\bf y}_1)
            \end{array},
        \end{eqnarray}
      which can be different from the decoder $\psi_1$ used in the case when no interference exists.

    \item[2)] The decoding function at User D is
        \begin{eqnarray}
            \begin{array}{cccl}
                \psi_2: & \mathbb{R}^{nN} & \rightarrow & \mathcal{M}_2 \\
                   & {\bf y}_2  &\mapsto & \hat{w}_2 = \psi_2({\bf y}_2)
            \end{array}.
        \end{eqnarray}
\end{enumerate}

%{\em \underline{Error Performance Criteria}:}
%\begin{enumerate}
%    \item[1)] For primary users, we define random variables
%        \begin{equation}\label{eqnErrorseq2}
%        \tilde{E}_t = \left\{\begin{array}{cc}
%                       0 & {\rm~if~} \tilde{U}_t= U_t \\
%                       1 & {\rm~if~} \tilde{U}_t \neq U_t
%                     \end{array}
%                \right. {\rm~for~} t \geq 1.
%        \end{equation}
%        The performance of the link between User A and B is measured by $p\texttt{-}\limsup_{N \rightarrow \infty} \tilde{\Theta}_N$, where
%        \begin{equation}\label{eqnDefTheta2}
%            \tilde{\Theta}_N = \frac{\Sigma_{t=1}^N \tilde{E}_t}{N}, {\rm~for~} N = 1,2,\ldots.
%        \end{equation}
%
%    \item[2)] For secondary users, we define $\lambda_2^{(N)} = {\rm Pr}\{\hat{W}_2 \neq W_2\}$.
%\end{enumerate}

{\em \underline{Error Performance Criteria}:}
\begin{enumerate}
    \item[1)] For the decoding at User~B, we define random variables
        \begin{equation}\label{eqnErrorseq2}
        \tilde{E}_t = \left\{\begin{array}{cc}
                       0, & {\rm~if~} \tilde{U}_t= U_t \\
                       1, & {\rm~if~} \tilde{U}_t \neq U_t
                     \end{array}
                \right. {\rm~for~} t \geq 1.
        \end{equation}
        The performance of the decoding is measured by $p\texttt{-}\limsup \tilde{\Theta}_N$, where
        \begin{equation}\label{eqnDefTheta2}
            \tilde{\Theta}_N = \frac{\Sigma_{t=1}^N \tilde{E}_t}{N}, {\rm~for~} N = 1,2,\ldots.
        \end{equation}

    \item[2)] For the decoding at User~D, we use as usual the WEP $\lambda_2^{(N)} = {\rm Pr}\{\hat{W}_2 \neq W_2\}$ to measure the performance.
\end{enumerate}

We make an assumption that User C knows exactly the coding function $\phi_1$ and attempts to find the optimal coding function $\phi_2$ under certain constraints. We also assume that User B and D know exactly the coding functions and attempt to find optimal decoding functions under certain criteria.

\begin{definition}\label{DefAchR2}
A rate $R_2$ is {\em achievable} for the secondary users, if for any $\delta > 0$, there exists a sequence of coding/decoding functions $(\phi_2, \psi_2)$ of coding rates greater than or equal to $R_2 - \delta$ such that $\lim_{N\rightarrow \infty} \lambda_2^{(N)} = 0$.
\hfill \ding{113}
\end{definition}

%\begin{definition}\label{DefAccR2}
%A rate $R_2$ is {\em accessible} for the secondary users if $R_2$ is achievable and there exists a decoder $\tilde{\psi}_1$ at the primary receive such that the GTC $\mathscr{C}_1$ is $\varepsilon$-satisfactory, that is, $p\texttt{-}\limsup_{N \rightarrow \infty} \tilde{\Theta}_N \leq \varepsilon$.
%\hfill \ding{113}
%\end{definition}
%
%\begin{definition}\label{DefAccC2}
%The {\em accessible capacity} for the secondary users is defined as the supremum of all achievable rates, i.e.,
%\begin{equation}
%    C_a = \sup \{R_2: R_2~{\rm is~accessible}\}.
%\end{equation}
%\vskip -0.3cm
%\hfill \ding{113}
%\end{definition}

\begin{definition}\label{DefAccR2}
A rate $R_2$ is {\em accessible} for the secondary users if $R_2$ is achievable and there exists a decoder $\tilde{\psi}_1$ at the primary receiver such that the GTC $\mathscr{C}_1$ is $\varepsilon$-satisfactory, that is, $p\texttt{-}\limsup \tilde{\Theta}_N \leq \varepsilon$.
\hfill \ding{113}
\end{definition}

\begin{definition}\label{DefAccC2}
The {\em accessible capacity} for the secondary users is defined as the supremum of all accessible rates, i.e.,
\begin{equation}
    C_a(\mathscr{C}_1, \varepsilon) = \sup \{R_2: R_2~{\rm is~accessible}\}.
\end{equation}
\vskip -0.3cm
\hfill \ding{113}
\end{definition}

{\bf Remark.} The notation $C_a(\mathscr{C}_1, \varepsilon)$ indicates that the accessible capacity depends on the GTC $\mathscr{C}_1$ and the error performance requirement $\varepsilon$ by the primary users. When no confusion arises in the context, we use notation $C_a$ instead for ease.

{\bf Problem formulation:} The problem is formulated as,
\begin{center}
   {\em given $\mathscr{C}_1$ and $\varepsilon$, find $C_a$.}
\end{center}

\subsection{Relationships Between Our Formulation And Existing Works}

\begin{enumerate}
  \item As we have mentioned in Introduction, the problem of finding the {\em capacity region}\footnote{The explicit definition of the capacity region may be found in the references, say~\cite{Carleial78}.} of the classical GIFC is equivalent to that of determining whether or not a pair of codes $(\mathscr{C}_1, \mathscr{C}_2)$ exist with any given respective coding rates $(R_1,R_2)$ such that the decoding error probabilities are arbitrarily small. To determine the capacity region, both $\mathscr{C}_1$ and $\mathscr{C}_2$ are allowed to be varied for the purpose of optimization. It is a two-dimensional optimization problem. In our formulation, the primary encoder $\mathscr{C}_1$ is assumed to be fixed and only the secondary encoder $\mathscr{C}_2$ is allowed to be varied for the purpose of finding the accessible capacity $C_a$. This is a one-dimensional optimization problem, which could be more tractable than the two-dimensional one. However, the accessible capacity is closely related to the capacity region as illustrated in the following.
        \begin{itemize}
          \item For large $\varepsilon$, the pair $(R_1, C_a)$ may fall outside the capacity region of the GIFC.

          \item When $\varepsilon \rightarrow 0$, the pair $(R_1, C_a)$ must fall inside the capacity region.

          \item Specifically, the pair $(R_1, C_a^*)$ must lie on the boundary of the capacity region, where $C_a^*$ is defined as
               \begin{equation}
                    C_a^* = \lim_{\varepsilon \rightarrow \infty} \sup_{\left\{\mathscr{C}_1\right\}} \{C_a(\mathscr{C}_1, \varepsilon)\}
               \end{equation}
               and the supremum is taken over all possible $\varepsilon$-satisfactory GTCs $\mathscr{C}_1$ of rate $R_1$.
          %
          %\item Specifically, the pair $(R_1, C_a^*)$ must lie on the boundary of the capacity region, where $C_a^*$ is defined as
%               \begin{equation}
%                    C_a^* = \lim_{\varepsilon \rightarrow 0} \sup \{C_a(\mathscr{C}_1, \varepsilon)\}
%               \end{equation}
%               and the supremum is taken over all codes of rate $R_1$.
        \end{itemize}
  \item In cognitive GIFCs~\cite{Devroye06}, the secondary encoder is assumed to have knowledge of the primary message. However, in our formulation, to design the secondary encoder, it needs only the structure of the primary GTC along with the primary error performance requirement $\varepsilon$ and the primary transmission power $P_1$.
      In~\cite{Jovicic09}, Jovi\v{c}i\'{c} and Viswanath presented a model that imposes two imperative constraints~(referred to as {\em coexistence conditions}) on the secondary users~(with noncausal knowledge of the primary message): 1)~it creates no rate degradation for the primary user in its vicinity, and 2)~the primary receiver uses a single-user decoder,
      %\footnote{Actually, in~\cite{Jovicic09}, for the cognitive system with ``low interference'' at the primary receiver, the single-user decoder at the primary user was shown to be optimal in the viewpoint of the cognitive capacity, while for the cognitive system with ``high interference'' at the primary receiver, the multiuser decoder at the primary user was shown to be optimal in the viewpoint of the capacity region.},
      just as it would in the absence of the cognitive radio.
      In our formulation, the first constraint is strengthened by fixing the primary encoder, and the second constraint is relaxed by allowing the primary decoder to be changed. Since the primary rate, the primary power and the channel coefficients are fixed in our formulation, OIC~\cite{Popovski07} is not applicable and we do not consider outage performance~\cite{DiTaranto11} in this paper.
  \item In most existing works on GIFCs, both the primary users and the secondary users~(with or without cooperation) are assumed to use random Gaussian codebooks with an exception of~\cite{Moshksar11}. In~\cite{Moshksar11}, the primary transmitter selects randomly~(possibly from a very high dimensional space) a fixed-size constellation for each transmission frame and then reveals the constellation non-causally to both primary users and secondary users, while the secondary transmitter is assumed to use a random Gaussian codebook. In our formulation, the primary transmitter uses not only a fixed constellation but also a fixed code structure. Our formulation is also closely related  to~\cite{Baccelli11}. Actually, in our formulation, the primary transmitter uses a point-to-point code, while the secondary transmitter uses a multi-user code. That is, the secondary code must be designed considering the primary point-to-point code. Hence, a good point-to-point code may not be a good candidate for a secondary code.
\end{enumerate}

\section{Bounds on the Accessible Capacity}\label{SecBoundCap}

In this section, we derive bounds on the accessible capacity for the secondary users. For doing so,
we rewrite the considered system in~(\ref{eqnIFCseq}) in  terms of random variables as
\begin{equation}\label{eqnIFCseqRV}
           \begin{array}{ccrcrcl}
            {\bf Y}_1 & = & {\bf X}_1 & + & a_{21} {\bf X}_2 & + & {\bf Z}_1\\
            {\bf Y}_2 & = & a_{12} {\bf X}_1 & + & {\bf X}_2 & + & {\bf Z}_2
           \end{array},
\end{equation}
where
\begin{itemize}
  \item ${\bf X}_1$ is a random sequence of length $nN$ whose pmf $p({\bf x}_1)$ can be determined by the i.u.d. input to the encoder together with the initial state of the encoder, that is, ${\bf X}_1$ is uniformly distributed over $\mathscr{C}_1$ for a given initial state;
  \item ${\bf Z}_1$ and ${\bf Z}_2$ are two sequences of i.i.d. Gaussian random variables of variance one;
  \item ${\bf X}_2$ is a random sequence of length $nN$ whose distribution is to be determined.
  %\item ${\bf X}_2$ is a random sequence of length $nN$, \textcolor{blue}{which is to be determined}.
\end{itemize}

From the previous section, the accessible capacity depends on the following parameters: the GTC structure at the primary transmitter, the primary transmission power $P_1$, the error performance requirement $\varepsilon$ by the primary users, and the transmission power constraint $P_2$ at the secondary transmitter. Given the structure of the GTC, we may treat $P_1$ and $\varepsilon$ as a single parameter by removing the so-called {\em interference margin} as defined below, which is essentially the same as that defined in~\cite{Devroye11}.

Let $\varepsilon$ be the error performance requirement by the primary users. Obviously, $p\texttt{-}\limsup \Theta_N$ as defined in~(\ref{pLimSupErr}) depends not only on the GTC but also on the power $P_1$. Given the GTC, define
\begin{equation}\label{EqnDefP1*}
    P_1^* \stackrel{\Delta}{=}\inf \{P|P~{\rm is~the~primary~transmission~power~such~that}~p\texttt{-}\limsup \Theta_N \leq \varepsilon\}.
\end{equation}
Similar to~\cite{Devroye11}, we call $\delta_1 \stackrel{\Delta}{=} P_1/P_1^*$ {\em interference margin}. Obviously, $\delta_1 \geq 1$ since we assume that the error performance is satisfied for the interference-free AWGN channel in the beginning.

%\subsection{Removal of Interference Margin~(i.e., $\delta_1 = 1)$}\label{SubsecTriIM}
\subsection{Removal of Interference Margin}\label{SubsecTriIM}

Temporarily, we assume that the interference margin is removed by the primary transmitter, i.e., $\delta_1 = 1$. This assumption means that the primary transmitter works at the most ``economic'' power level $P_1^*$, and (hence) is reasonable since the GTC was originally designed for the primary users in the absence of the interference. For the derivations in this subsection, we assume that ${\bf X}_2$ is a sequence of discrete random variables, whose pmf is denoted by $p({\bf x}_2)$ for ${\bf x}_2 \in \mathcal{X}_2^N$, where $\mathcal{X}_2 \subseteq \mathbb{R}^n$ is a finite set. This assumption is reasonable since we are primarily interested in the low SNR regime.

%Noting that the performance loss caused by constraining the input to be selected from a finite constellation can be made negligible by making the quantization sufficiently fine~\cite[Chapters~2.4,~2.5 and~7]{Gallager68}.
%~\cite{Cover91}\footnote{For completeness, we show the detail in Appendix~\ref{AppendixFSCarbitraryEnsbl}.}.
%~\cite[Chapters~2.4,~2.5 and~7]{Gallager68}\footnote{In~\cite[Chapters~2.4,~2.5 and~7]{Gallager68}, the discrete-time memoryless channel with arbitrary input and output ensembles was transformed into its reduced discrete memoryless channel by using the partitions of the input ensemble and the output ensemble. In such a way, the mutual information and even the coding theorem of the original memoryless channel were investigated. Similar approaches can be extended for the discrete-time finite state channel with continuous/arbitrary ensembles, which is given in Appendix~\ref{AppendixFSCarbitraryEnsbl}.}.

Considering the GTC structure at the primary transmitter, we have two finite state channels~(FSCs) as shown in the following lemma.

%Consider the case of $\delta_1 = 1$, which means that the primary transmitter works at the most ``economic" power level $P_1^*$, and (hence) is reasonable since the GTC was originally designed for the primary users in the absence of the interference. That is, there does not exist interference margin at the primary receiver. \textcolor{blue}{In this case, we assume that $\mathcal{X}_2 \subseteq \mathbb{R}^n$ is a finite set. Let ${\bf X}_2$ be a sequence of discrete random variables over $\mathcal{X}_2^{N}$, whose pmf is denoted by $p({\bf x}_2)$. Let $\mathscr{C}_2$ be the code used by the secondary transmitter.}
%
%\textcolor{blue}{Considering the GTC structure at the primary transmitter, we have two finite state channels~(FSCs) as shown in the following lemma.}

\begin{lemma}\label{LMFSC}
Both the links ${\bf X}_2\rightarrow {\bf Y}_2$ and ${\bf X}_2\rightarrow {\bf Y}_1$ can be viewed as FSCs where the processes of channel states are Markovian.
\hfill \ding{113}
\end{lemma}

\begin{IEEEproof}
From ${\bf Y}_2 = a_{12} {\bf X}_1 + {\bf X}_2 + {\bf Z}_2$ in~(\ref{eqnIFCseqRV}), we can see that the transition probability from ${\bf x}_2$ to ${\bf y}_2$ depends on the random sequence ${\bf x}_1$, which corresponds to a path through the trellis of the GTC $\mathscr{C}_1$. To prove that this channel is an FSC as defined in~\cite{Gallager68}, we need to define a channel state such that, conditioned on the channel state and the current input, the channel output is statistically independent of all previous inputs and outputs. Let $s_t\in \mathcal{S}$ be the trellis state at the $t$-th stage of the GTC. We can see that the channel ${\bf X}_2\rightarrow {\bf Y}_2$ is then {\em completely characterized} by
\begin{equation}\label{eqnFSCTrans}
    f(y_{2,t}, s_{t} | x_{2,t}, s_{t-1}) = \sum \frac{1}{M}
    \frac{1}{(2\pi)^{n/2}} \exp\left\{-\frac{\|y_{2,t}-a_{12}x_{1,t}-x_{2,t}\|^2}{2}\right\},
\end{equation}
where $x_{1,t} = \sqrt{P_1}c(b)$ and the summation is over all branches $b$ connecting $s_{t-1}$ and $s_{t}$. This  allows us to follow Gallager~\cite{Gallager68} and to work with the conditional probability $f({\bf y}_2, s_N|{\bf x}_2, s_0)$, which can be calculated inductively from
\begin{equation}\label{eqnFSCSeqTrans}
    f({\bf y}_2, s_N|{\bf x}_2, s_0) = \sum\limits_{s_{N-1}} f(y_{2,N}, s_{N} | x_{2,N}, s_{N-1})
    f({\bf y}_2^{(N-1)}, s_{N-1}|{\bf x}_2^{(N-1)}, s_0),
\end{equation}
where ${\bf x}_2^{(N-1)} = (x_{2, 1}, x_{2, 2}, \cdots, {x_{2, N-1}})$ and ${\bf y}_2^{(N-1)} = (y_{2, 1}, y_{2, 2}, \cdots, {y_{2, N-1}})$. The final state can be summed over to give

\begin{equation}\label{eqnFSCSeqTrans1}
    f({\bf y}_2|{\bf x}_2, s_0) = \sum\limits_{s_{N}}f({\bf y}_2, s_{N}|{\bf x}_2, s_0).
\end{equation}

The channel state process $\{S_t\}$ is a Markov process, which evolves freely~(independently from the input ${\bf x}_2$) with the following transition probabilities
 \begin{equation}\label{eqnStateTrans}
    p_{S_t|S_{t-1}}(s_t|s_{t-1}) = \frac{~number~of~branches~connecting~s_{t-1}~and~s_t}{M}
 \end{equation}
for $s_{t-1} \in \mathcal{S}, s_{t} \in \mathcal{S}$.
Therefore the link ${\bf X}_2\rightarrow {\bf Y}_2$ is a {\em noncontrollable} FSC~\cite{Vontobel08,Huang09,Huang12}.
Similarly, the link ${\bf X}_2 \rightarrow {\bf Y}_1$ is also a noncontrollable FSC with the channel state process $\{S_t\}$.
\end{IEEEproof}

Following Gallager~\cite{Gallager68}, define
\begin{equation}\label{eqnDefUppBound}
    C_U^{(N)} \stackrel{\Delta}{=} \frac{1}{nN} \max_{\{p({\bf x}_2)\}} \max_{s_0} I({\bf X}_2; {\bf Y}_2 | s_0)
\end{equation}
and
\begin{equation}\label{eqnDefLowBound}
    C_L^{(N)} \stackrel{\Delta}{=} \frac{1}{nN} \max_{\{p({\bf x}_2)\}}\min_{s_0} \min\{I({\bf X}_2; {\bf Y}_1 | s_0), I({\bf X}_2; {\bf Y}_2 | s_0)\},
\end{equation}
where the set $\{p({\bf x}_2)\}$ consists of all possible pmfs $p({\bf x}_2)$ over $\mathcal{X}_2^N$ such that ${\bf E}\left[\|{\bf X}_2\|^2\right] \leq nNP_2$.

\begin{lemma}\label{LMlimits}
The limits $C_U = \lim\limits_{N\rightarrow \infty} C_U^{(N)}$ and $C_L = \lim\limits_{N\rightarrow \infty} C_L^{(N)}$ exist.
\hfill \ding{113}
\end{lemma}

\begin{IEEEproof}
The existence of $C_U$ can be proved by applying directly Theorem 4.6.1 in~\cite{Gallager68}\footnote{For completeness, Theorem 4.6.1 in~\cite{Gallager68} is rephrased in Appendix.}. To prove the existence of $C_L$, we define a new FSC whose channel states are drawn from $\bar{\mathcal{S}} \stackrel{\Delta}{=} \{1,2\} \times \mathcal{S}$. Given an initial state $\bar{s}_0 = (i, s_0)$, the new FSC channel
is completely characterized by
\begin{equation}\label{eqnNewFSC}
    f_{{\bf Y}|{\bf X}_2}({\bf y}|{\bf x}_2, \bar{s}_0) = \left\{
                                                \begin{array}{cc}
                                                 f_{{\bf Y}_1|{\bf X}_2}({\bf y}|{\bf x}_2, s_0),  & {\rm if} ~~ i = 1 \\
                                                      f_{{\bf Y}_2 | {\bf X}_2}({\bf y}|{\bf x}_2, s_0), & {\rm if} ~~ i = 2
                                                \end{array}
                                                \right..
\end{equation}
Then the definition of $C_L^{(N)}$ in~(\ref{eqnDefLowBound}) can be rewritten as
\begin{equation*}
    C_L^{(N)} \stackrel{\Delta}{=} \frac{1}{nN} \max_{\{p({\bf x}_2)\}}\min_{\bar{s}_0}I({\bf X}_2; {\bf Y}| \bar{s}_0)
\end{equation*}
and the existence of $C_L$ becomes obvious from Theorem 4.6.1 in~\cite{Gallager68}.
\end{IEEEproof}

%\begin{lemma}\label{LMlimits}
%The limits $C_U = \lim\limits_{N\rightarrow \infty} C_U^{(N)}$ and $C_L = \lim\limits_{N\rightarrow \infty} C_L^{(N)}$ exist.
%%\begin{equation}
%%C_U = \inf_{N} \left\{C_U^{(N)} + \frac{\log |\mathcal{S}|}{nN} \right\}~{\rm and}~
%%C_L = \sup_{N} \left\{C_L^{(N)} - \frac{1 + \log |\mathcal{S}|}{nN} \right\}
%%\end{equation}
%%where $\mathcal{S}$ is the state set of FSCs ${\bf X}_2\rightarrow {\bf Y}_2$ and ${\bf X}_2\rightarrow {\bf Y}_1$~(also the state set of the primary GTC).
%\hfill \ding{113}
%\end{lemma}
%
%\begin{IEEEproof}
%See Appendix.%~\ref{ProofLMlimits}.
%\end{IEEEproof}

%The main result of this subsection is stated in the following theorem.

\begin{theorem}\label{THCapBound}
The accessible capacity $C_a$ is bounded as
\begin{equation}\label{eqnTHbounds}
    C_L \leq C_a \leq C_U.
\end{equation}
\vskip -0.5cm
\hfill \ding{113}
\end{theorem}

\begin{IEEEproof}
Intuitively, these bounds can be explained as follows. The term $I({\bf X}_2; {\bf Y}_2 | s_0)$ specifies the achievable rate for the secondary link, hence it appears in both the upper bound $C_U$ derived from~(\ref{eqnDefUppBound}) and the lower bound $C_L$ derived from~(\ref{eqnDefLowBound}). The term $I({\bf X}_2; {\bf Y}_1 | s_0)$ specifies the achievable rate for the cross link ${\bf X}_2 \rightarrow {\bf Y}_1$. Hence, putting $I({\bf X}_2; {\bf Y}_1 | s_0)$ into the lower bound $C_L$ derived from~(\ref{eqnDefLowBound}) makes it possible for the primary receiver to cancel the interference from the secondary transmitter first before doing its own decoding. The formal proof is given below.

Firstly, we prove that any rate $R_2 > C_U$ is not achievable for the link ${\bf X}_2 \rightarrow {\bf Y}_2$.
It is equivalent to proving that, for any code, if the probability of decoding error $\lambda_2^{(N)} = {\rm Pr}\{\hat{W}_2 \neq W_2\} \rightarrow 0$ as $N \rightarrow \infty$, then the coding rate $R_2 \leq C_U$.

Actually, from Fano's inequality and data processing inequality~\cite{Cover91}, we have
\begin{eqnarray}\label{eqnFano1}
    nNR_2 = H(W_2|s_0)
    &=& H(W_2 | {\bf Y}_2, s_0) + I(W_2; {\bf Y}_2|s_0) \nonumber\\
    &\leq& 1 + \lambda_2^{(N)} n N R_2 + I(W_2; {\bf Y}_2|s_0)\nonumber\\
    &\leq& 1 + \lambda_2^{(N)} n N R_2 + I({\bf X}_2; {\bf Y}_2|s_0).
\end{eqnarray}
Dividing by $nN$,
\vskip -0.8cm
\begin{eqnarray}
    R_2 &\leq& \frac{1}{nN} + \lambda_2^{(N)} R_2 + \frac{1}{nN} I({\bf X}_2; {\bf Y}_2|s_0)\\
        &\leq& \frac{1}{nN} + \lambda_2^{(N)} R_2 + C_U^{(N)}.
\end{eqnarray}
As $N \rightarrow \infty$, we have $R_2 \leq C_U$ since $\frac{1}{nN} \rightarrow 0$ and $\lambda_2^{(N)} \rightarrow 0$.

Secondly, we prove that any rate $R_2 < C_L$ is accessible. It is equivalent to proving that $R_2 < C_L$ is achievable and the GTC $\mathscr{C}_1$ is still $\varepsilon$-satisfactory in the presence of the secondary transmission rate $R_2$.

%To prove the achievability of $R_2$, we apply Theorem~5.9.2 in~\cite{Gallager68} \textcolor{blue}{to the newly defined FSC~(\ref{eqnNewFSC}), which is introduced to prove the existence of $C_L$ in Appendix}. We have the following facts.

Applying Theorem 5.9.2 in~\cite{Gallager68} to the newly defined FSC in~(\ref{eqnNewFSC}), we have the following facts. For any $\varepsilon > 0$, there exists $N(\varepsilon)$ such that for each $N \geq N(\varepsilon)$ and each $R_2 \geq 0$ there exists a block code $\mathscr{C}_2$ with rate $R_2$ and codeword length $nN$ such that, for all initial states $\bar{s}_0 \in \bar{\mathcal{S}}$,
\vskip -0.8cm
\begin{equation}
    \begin{array}{ccc}
      \lambda_1^{(N)} & \leq & 2^{-N[E_{r}(R_2) - \varepsilon]}\\
      \lambda_2^{(N)} & \leq & 2^{-N[E_{r}(R_2) - \varepsilon]}
    \end{array},
\end{equation}
where
\begin{itemize}
  \item $\lambda_1^{(N)}$ is the average probability of erroneously decoding $W_2$ from the received sequence ${\bf Y}_1$ at User B by the maximum-likelihood decoding algorithm $\tilde{\psi}_{1,1}$;
      %, i.e., $\tilde{\psi}_{1,1}({\bf y}_1) = \tilde{w}_2 = \arg \max\limits_{w_2} f({\bf y}_1|{\bf x}_2(w_2))$ and $\lambda_1^{(N)} = {\rm Pr}\{\tilde{W}_2 \neq W_2\}$;
  \item $\lambda_2^{(N)}$ is the average probability of erroneously decoding $W_2$ from the received sequence ${\bf Y}_2$ at User D by the maximum-likelihood decoding algorithm\footnote{The maximum-likelihood decoding algorithm $\psi_{2}$ is defined as~\cite[Chapter~5.9]{Gallager68}
      \begin{equation*}
      \hat{w}_2 \stackrel{\Delta}{=} \psi_{2}({\bf y}_2) = \arg \max_{w_2} \sum_{s_0} \frac{1}{|\mathcal{S}|}f({\bf y}_2|{\bf x}_2(w_2),s_0)
      \end{equation*}
      where $f({\bf y}_2|{\bf x}_2(w_2),s_0)$ can be computed by combining~(\ref{eqnFSCTrans}),~(\ref{eqnFSCSeqTrans}) and~(\ref{eqnFSCSeqTrans1}). The maximum-likelihood decoding algorithm $\tilde{\psi}_{1,1}$ can be defined similarly.} $\psi_{2}$;
      %, i.e., $\psi_{2}({\bf y}_2) = \hat{w}_2 = \arg \max\limits_{w_2} f({\bf y}_2|{\bf x}_2(w_2))$ and $\lambda_2^{(N)} = {\rm Pr}\{\hat{W}_2 \neq W_2\}$;
  \item $E_{r}(R_2)$ is the {\em random coding error exponent}~\cite{Gallager68}, which is strictly positive for $R_2 < C_L$.
\end{itemize}
Therefore, as $N \rightarrow \infty$, $\lambda_1^{(N)} \rightarrow 0$ and $\lambda_2^{(N)} \rightarrow 0$. The latter implies that the rate $R_2 < C_L$ is achievable for the secondary link ${\bf X}_2 \rightarrow {\bf Y}_2$.

%To prove $\varepsilon$-satisfactoriness of $\mathscr{C}_1$ ,

To complete the proof, we need to find a decoder $\tilde{\psi}_1$ such that the error performance requirement $\varepsilon$ is still fulfilled, that is, $p\texttt{-}\limsup \tilde{\Theta}_N \leq \varepsilon$. Such decoders do exist, one of which is the following two-stage decoder.
\begin{description}
    \item[{\underline{Step~1}}:]~Upon receiving ${\bf y}_1$, User B utilizes the maximum-likelihood decoder $\tilde{\psi}_{1,1}$ to get an estimated message $\tilde{w}_2$. The probability of decoding error $\lambda_1^{(N)} = {\rm Pr}\{\tilde{W}_2 \neq W_2\}$ goes to zero as $N$ goes to infinity. For convenience, we introduce a random variable as
        \begin{equation}
            \Upsilon_1 = \left\{
            \begin{array}{cc}
                0, & \tilde{W}_2 = W_2\\
                1, & \tilde{W}_2 \neq W_2
            \end{array}
            \right..
        \end{equation}
        Then $\lambda_1^{(N)} = {\rm Pr} (\Upsilon_1 = 1)$.

    \item[{\underline{Step~2}}:]~User B re-encodes $\tilde{w}_2$ to get an estimated coded sequence ${\bf \tilde{x}}_2 = \phi_2(\tilde{w}_2)$.
        Then User B uses the primary decoder $\psi_1$ to decode the sequence $\tilde{\bf y}_1 = {\bf y}_1 - a_{21} {\bf \tilde{x}}_2$ to get a sequence of estimated messages $\tilde{w}_1 = (\tilde{u}_{1}, \tilde{u}_{2}, \ldots, \tilde{u}_{N})$.
\end{description}

The above two-stage decoder, denoted by $\tilde{\psi}_{1} = {\psi_{1}} \circ \tilde{\psi}_{1,1}$, is a successive {\em full} interference cancellation  decoding scheme. For such a two-stage decoder, the statistical dependence among the error random variables $\{\tilde{E}_t\}$ as defined in~(\ref{eqnErrorseq2}) becomes even more complicated. On one hand, the erroneously-decoding of $w_2$ at User $B$ may cause burst errors in $\tilde{w}_1$ at the second stage. On the other hand, the correctly-decoding of $w_2$ at User $B$ indicates that the link ${\bf X}_2 \rightarrow {\bf Y}_1$ is not that noisy, equivalently, that the sum of the transmitted codeword ${\bf x}_1$ and the Gaussian noise sequence ${\bf z}_1$ is not that ``strong''. This implies that, even in the case of $\Upsilon_1 = 0$, the random variables $\{\tilde{E}_t\}$ may still have a different distribution from the random variables $\{E_t\}$ as defined in~(\ref{eqnErrorseq}) for the interference-free AWGN channel. Fortunately, this complicatedness does not affect the $\varepsilon$-satisfactoriness of the GTC $\mathscr{C}_1$. Actually, for random variables $\tilde{\Theta}_N $ as defined in~(\ref{eqnDefTheta2}), we have
\begin{eqnarray}
    {\rm Pr}\{\tilde{\Theta}_N > \varepsilon\}
    &=& {\rm Pr}\{\tilde{\Theta}_N > \varepsilon, \Upsilon_1 = 0\} + {\rm Pr}\{\tilde{\Theta}_N > \varepsilon, \Upsilon_1 = 1\} \nonumber\\
    &\leq& {\rm Pr}\{\tilde{\Theta}_N > \varepsilon, \Upsilon_1 = 0\} + \lambda_1^{(N)} \nonumber\\
    &=& {\rm Pr}\{\Theta_N > \varepsilon, \Upsilon_1 = 0\} + \lambda_1^{(N)} \nonumber\\
    &\leq& {\rm Pr}\{\Theta_N > \varepsilon\} + \lambda_1^{(N)}.
\end{eqnarray}
Since the GTC $\mathscr{C}_1$ is $\varepsilon$-satisfactory under the decoder $\psi_{1}$~(implying $\lim\limits_{N\rightarrow \infty}{\rm Pr}\{\Theta_N > \varepsilon\} = 0$) and $R_2 < C_L$~(implying $\lim\limits_{N\rightarrow \infty}\lambda_1^{(N)} = 0$), we have $\lim\limits_{N\rightarrow \infty}{\rm Pr}\{\tilde{\Theta}_N > \varepsilon\} = 0$, which is equivalent to $p\texttt{-}\limsup\tilde{\Theta}_N \leq \varepsilon$.
\end{IEEEproof}

%From the proof of Lemmas~\ref{LMFSC} and~\ref{LMlimits}, we see that, given the GTC structure of $\mathscr{C}_1$, both the secondary link ${\bf X}_2 \rightarrow {\bf Y}_2$ and the cross link ${\bf X}_2 \rightarrow {\bf Y}_2$ from the secondary transmitter to the primary receiver are uniquely determined by the transmission powers $P_1$ and $P_2$. Hence, at a first glance, it seems that the derived upper and lower bounds $C_U$ and $C_L$ are not directly related to the error performance requirement $\varepsilon$. However, from the proof of Theorem~\ref{THCapBound}, we can see that...... Actually, $P_1$ plays the same role as $\varepsilon$ does,.....
%
%Therefore, both the upper and lower bounds $C_U$ and $C_L$ are not directly related to the error performance requirement $\varepsilon$.

%An immediate consequence of Theorem~\ref{THCapBound} is the following corollary, which is related to a similar case when interferences are strong.
%
%\begin{corollary}\label{CorStrIF}
%If $I({\bf X_2}; {\bf Y_2}|s_0) \leq I({\bf X_2}; {\bf Y_1}|s_0)$ holds for all pmfs $p({\bf x}_2)$ and all initial states $s_0\in \mathcal{S}$, then $C_a = C_L = C_U$.
%\hfill \ding{113}
%\end{corollary}

An immediate consequence of Theorem~\ref{THCapBound} is the following corollary, which is related to the case of strong interference at the primary receiver.

\begin{corollary}\label{CorStrIF}
For the new formulation of the GIFC, if the strong interference condition
\begin{equation}\label{EqnStngCond}
    I({\bf X_2}; {\bf Y_2}|s_0) \leq I({\bf X_2}; {\bf Y_1}|s_0)
\end{equation}
holds for all pmfs $p({\bf x}_2)$ and all initial states $s_0\in \mathcal{S}$, then the upper bound and the lower bound coincide and the accessible capacity is $C_a = C_L = C_U$.
\hfill \ding{113}
\end{corollary}

%Although the one-dimensional accessible capacity of the new formulation is different from the two-dimensional capacity region of the cognitive IFC, the strong interference condition in~(\ref{EqnStngCond}) in Corollary~\ref{CorStrIF} is analogous to the case investigated in~\cite[Theorem~5]{Maric07}, where the capacity region for certain strong interference cognitive IFCs was presented.

The condition in~(\ref{EqnStngCond}) means that the cross link from the secondary transmitter to the primary receiver is better/stronger than the secondary link~(say, $a_{21}^2 > 1$ and $a_{12}^2 < 1$).
If this is the case, any rate that is achievable through the secondary link is also achievable through the cross link from the secondary transmitter to the primary receiver, which admits a successive interference cancellation decoding algorithm at the primary receiver.
Also noticing that the strong interference condition~(\ref{EqnStngCond}) is analogous to the case investigated in~\cite[Theorem~5]{Maric07}, where the capacity region for a certain strong interference cognitive IFC was presented.

%As a special class of GTCs, block codes satisfy the property that all components $x_{1,t}\in \mathbb{R}^n$ in the coded sequence ${\bf x}_1 = (x_{1,1}, x_{1,2}, \ldots, x_{1,N})$ are independent. In this case, both the links ${\bf X}_2 \rightarrow {\bf Y}_1$ and ${\bf X}_2 \rightarrow {\bf Y}_2$ can be viewed as {\em block-wise} memoryless channels, hence Theorem~\ref{THCapBound} and Corollary~\ref{CorStrIF} can be simplified.

As a special class of GTCs, block codes satisfy the property that all components $x_{1,t}\in \mathbb{R}^n$ in the coded sequence ${\bf x}_1 = (x_{1,1}, x_{1,2}, \ldots, x_{1,N})$ are independent since the GTC of a block code has only one state and each $n$-dimensional coded signal $x_{1,t}$ is driven by the message $u_t$ (see the description of the GTC and Examples~\ref{ExRCBPSK} and~\ref{ExEHCBPSK} in Sec.~\ref{SubSecPL}). In this case, both the links ${\bf X}_2 \rightarrow {\bf Y}_1$ and ${\bf X}_2 \rightarrow {\bf Y}_2$ can be viewed as {\em block-wise} memoryless channels.
%\footnote{The difference between the block-wise memoryless channel here and the ``usual'' memoryless channel is that the channel input signal $X_2$ is a ``super'' $n$-dimensional signal rather than a ``usual'' one-dimensional signal, where $n$ is the length of the block GTC at the primary user.},
That is, the channel laws of the links ${\bf X}_2 \rightarrow {\bf Y}_1$ and ${\bf X}_2 \rightarrow {\bf Y}_2$ can be characterized by, for all ${\bf x}_2, {\bf y}_1, {\bf y}_2 \in \mathbb{R}^{nN}$,
\begin{equation}
f\left(\left.{\bf y}_1\right|{\bf x}_2\right) = \prod_{t=1}^N f\left(\left. y_{1,t}\right|x_{2,t}\right)~{\rm and}~
f\left(\left.{\bf y}_2\right|{\bf x}_2\right) = \prod_{t=1}^N f\left(\left. y_{2,t}\right|x_{2,t}\right),
\end{equation}
respectively. Hence, Theorem~\ref{THCapBound} and Corollary~\ref{CorStrIF} can be simplified as Corollary~\ref{CorBlockGTC} and Corollary~\ref{CorBlockGTCStrong} as below, respectively.

\begin{corollary}\label{CorBlockGTC}
For conventional block codes~(GTCs with only one state), the bounds on the accessible capacity are reduced to
\begin{equation}\label{eqnDefUBblock}
    C_U = \frac{1}{n} \max_{\{p(x_2)\}} I(X_2; Y_2),
\end{equation}
and
\begin{equation}\label{eqnDefLBblock}
    C_L = \frac{1}{n} \max_{\{p(x_2)\}} \min\{I(X_2; Y_1), I(X_2; Y_2)\},
\end{equation}
where the set $\{p(x_2)\}$ consists of all possible pmfs over $\mathcal{X}_2 \!\subseteq\! \mathbb{R}^n$ such that ${\bf E}\left[\|X_2\|^2\right] \!\leq\! nP_2$.
\hfill \ding{113}
\end{corollary}

\begin{IEEEproof}
Since both the links ${\bf X}_2 \rightarrow {\bf Y}_1$  and ${\bf X}_2 \rightarrow {\bf Y}_2$ are block-wise memoryless with only one channel state, we can remove the initial state $s_0$ from the definitions of $C_U^{(N)}$ in~(\ref{eqnDefUppBound}) and $C_L^{(N)}$ in~(\ref{eqnDefLowBound}).

On one hand, for any pmf $p(x_2)$, if we take an i.i.d. sequence ${\bf X}_2$ (where $X_{2,t}\sim p(x_2)$) as the input process, we have
\begin{equation}\label{eqnMIRUBseq-single1}
I({\bf X}_2; {\bf Y}_2) = \sum_{t=1}^N I\!\left(X_{2,t};Y_{2,t}\right) = N I\!\left(\!X_{2};Y_{2}\right),
\end{equation}
and
\begin{equation}\label{eqnMIRUBseq-single2}
I({\bf X}_2; {\bf Y}_1) = \sum_{t=1}^N I\!\left(X_{2,t};Y_{1,t}\right) = N I\!\left(\!X_{2};Y_{1}\right).
\end{equation}
This implies $C_U^{(N)} \geq \frac{1}{n} I(X_2; Y_2)$ and $C_L^{(N)} \geq \frac{1}{n} \min\left\{I(X_2; Y_1),I(X_2; Y_2)\right\}$ for any i.i.d. input sequence. Hence,
\begin{equation}\label{eqnCapUBblock1}
    C_U^{(N)} \geq \frac{1}{n} \max_{\{p(x_2)\}} I(X_2; Y_2),
\end{equation}
and
\begin{equation}\label{eqnCapLBblock1}
    C_L^{(N)} \geq \frac{1}{n} \max_{\{p(x_2)\}} \min\left\{I(X_2; Y_1),I(X_2; Y_2)\right\}.
\end{equation}

On the other hand, for any {\em joint} pmf $p({\bf x}_2)$, using the same method as given in the proof of Lemma~8.9.2 in~\cite{Cover91}, we can prove that
\begin{equation}\label{eqnCapUBblock2}
    \frac{1}{nN} I({\bf X}_2; {\bf Y}_2) \leq \frac{1}{n} \max_{\{p(x_2)\}} I(X_2; Y_2).
\end{equation}
Combining~(\ref{eqnCapUBblock1}) and~(\ref{eqnCapUBblock2}), we have
\begin{equation}\label{eqnDefUBblock3}
    C_U = \frac{1}{n} \max_{\{p(x_2)\}} I(X_2; Y_2).
\end{equation}
Also from the proof of Lemma~8.9.2 in~\cite{Cover91}, we know that
\begin{eqnarray}\label{eqnCapLBblock2}
\frac{1}{N} I({\bf X}_2; {\bf Y}_2)
&\leq& \frac{1}{N} \sum_{t=1}^N I\!\left(X_{2,t};Y_{2,t}\right) \nonumber\\
&\stackrel{\rm (a)}{\leq}&  I\!\left(X_{2};Y_{2}\right)|_{X_2\sim {\bar Q}_2},
\end{eqnarray}
where ${\bar Q}_2$ is the pmf of $X_2$ defined as ${\bar p}(x_2) = \frac{1}{N} \sum_{t=1}^N p(X_{2,t}=x_2)$ and $p(X_{2,t}=x_2)$ is derived from the joint pmf $p({\bf x}_2)$, and inequality (a) results from the concavity of the mutual information $I(X_2,Y_2)$ with respect to the pmf $p(x_2)$ of $X_2$. Similarly, we have
\begin{equation}\label{eqnCapLBblock3}
\frac{1}{N} I({\bf X}_2; {\bf Y}_1)
\leq  I\!\left(X_{2};Y_{1}\right)|_{X_2\sim {\bar Q}_2}.
\end{equation}
From~(\ref{eqnCapLBblock2}) and~(\ref{eqnCapLBblock3}), we have
\begin{equation}\label{eqnCapLBblock4}
    \frac{1}{nN} \min \left\{I({\bf X}_2; {\bf Y}_1),I({\bf X}_2; {\bf Y}_2)\right\} \leq
    \frac{1}{n} \min \left\{I(X_2; Y_1)|_{X_2\sim {\bar Q}_2}, I(X_2; Y_2)|_{X_2\sim {\bar Q}_2}\right\},
\end{equation}
which implies that
\begin{equation}\label{eqnCapLBblock5}
    \frac{1}{nN} \min \left\{I({\bf X}_2; {\bf Y}_1),I({\bf X}_2; {\bf Y}_2)\right\} \leq
    \frac{1}{n} \max_{\{p(x_2)\}} \min \left\{I(X_2; Y_1),I(X_2; Y_2)\right\}.
\end{equation}
Combining~(\ref{eqnCapLBblock1}) and~(\ref{eqnCapLBblock5}), we have
\begin{equation}\label{eqnDefLBblock6}
    C_L = \frac{1}{n} \max_{\{p(x_2)\}} \min \left\{I(X_2; Y_1),I(X_2; Y_2)\right\}.
\end{equation}
\end{IEEEproof}

\begin{corollary}\label{CorBlockGTCStrong}
For conventional block codes~(GTCs with only one state), if $I(X_2; Y_2) \leq I(X_2; Y_1)$ holds for the pmf
$Q_2^* = \arg \max\limits_{\{p(x_2)\}} I(X_2; Y_2)$, then $C_a = C_L = C_U$.  That is,
\begin{equation}
     C_a = \left.\frac{1}{n} I(X_2; Y_2)\right|_{X_2\sim Q_2^*}.
\end{equation}
\hfill \ding{113}
\end{corollary}

\begin{IEEEproof}
Since $Q_2^*$ is the pmf of $X_2$ such that $Q_2^* = \arg \max\limits_{\{p(x_2)\}} I(X_2; Y_2)$, from the expression of $C_U$ in~(\ref{eqnDefUBblock}) in Corollary~\ref{CorBlockGTC}, we have $C_U=\left.\frac{1}{n} I(X_2; Y_2)\right|_{X_2\sim Q_2^*}$. First, by Theorem~\ref{THCapBound}, we have $C_L \leq C_U$. Second, if $\left.I(X_2; Y_2)\right|_{X_2\sim Q_2^*} \leq \left.I(X_2; Y_1)\right|_{X_2\sim Q_2^*}$, from the expression of $C_L$ in~(\ref{eqnDefLBblock}) in Corollary~\ref{CorBlockGTC}, we have
%\begin{equation}
%\frac{1}{n} \min \left\{\left.I(X_2; Y_2)\right|_{X_2\sim Q_2^*}, \left.I(X_2; Y_1)\right|_{X_2\sim Q_2^*}\right\}
% = \frac{1}{n} \left.I(X_2; Y_2)\right|_{X_2\sim Q_2^*} = C_U.
%\end{equation}
%Then from the definition of $C_L$ in~(\ref{eqnDefLBblock}), we have
\begin{eqnarray}
C_L &=& \frac{1}{n} \max_{\{p(x_2)\}} \min \left\{I(X_2; Y_2), I(X_2; Y_1)\right\} \nonumber\\
    &\geq& \frac{1}{n} \min \left\{\left.I(X_2; Y_2)\right|_{X_2\sim Q_2^*},
        \left.I(X_2; Y_1)\right|_{X_2\sim Q_2^*}\right\}\nonumber\\
    &=& \frac{1}{n} \left.I(X_2; Y_2)\right|_{X_2\sim Q_2^*}\nonumber\\
    &=& C_U.
\end{eqnarray}
%That is, $C_L \geq \left.\frac{1}{n} I(X_2; Y_2)\right|_{X_2\sim Q_2^*}=C_U$.
Therefore, the lower and upper bounds coincide. That is, $C_L=C_U=C_a=\left.\frac{1}{n} I(X_2; Y_2)\right|_{X_2\sim Q_2^*}$.
%First, we can see that, for those pmfs $\mathcal{P}_1=\{p(x_2)\}$ such that $I(X_2; Y_1) \leq I(X_2; Y_2)$,
%\begin{equation}
%\max_{\mathcal{P}_1} \min\{I(X_2; Y_1), I(X_2; Y_2)\} = \max_{\mathcal{P}_1} I(X_2; Y_1) \leq \max_{\mathcal{P}_1} I(X_2; Y_2).
%\end{equation}
%We can also see that, for those pmfs $\mathcal{P}_2=\{p(x_2)\}$ such that $I(X_2; Y_1) > I(X_2; Y_2)$,
%\begin{equation}
%\max_{\mathcal{P}_2} \min\{I(X_2; Y_1), I(X_2; Y_2)\} = \max_{\mathcal{P}_2} I(X_2; Y_2).
%\end{equation}
%Hence, from the definition of $C_L$ in~(\ref{eqnDefLBblock}) and the property that the set of all possible pmfs satisfies $\{p(x_2)\}=\mathcal{P}_1\bigcup\mathcal{P}_2$, we have
%\begin{equation}
%C_L \leq \frac{1}{n} \max\limits_{\{p(x_2)\}} I(X_2;Y_2) = \left.\frac{1}{n} I(X_2; Y_2)\right|_{X_2\sim Q_2^*}, {~\rm where~}Q_2^* = \arg \max\limits_{\{p(x_2)\}} I(X_2; Y_2).
%\end{equation}
%
%Second, if $I(X_2; Y_2) \leq I(X_2; Y_1)$ at the pmf $Q_2^*$, then $\min \left\{I(X_2; Y_1), I(X_2; Y_2)\right\} = I(X_2;Y_2)$ at the pmf $Q_2^*$. This implies that $C_L \geq \left.\frac{1}{n} I(X_2; Y_2)\right|_{X_2\sim Q_2^*}$. Therefore, $C_L = \left.\frac{1}{n} I(X_2; Y_2)\right|_{X_2\sim Q_2^*}$. Finally, from the definition of the lower bound $C_U$ in~(\ref{eqnDefUBblock}), we have $C_L=C_U=C_a=\left.\frac{1}{n} I(X_2; Y_2)\right|_{X_2\sim Q_2^*}$.
\end{IEEEproof}

Note that the condition in Corollary~\ref{CorBlockGTCStrong} is even slightly weaker than that in Corollary~\ref{CorStrIF}.

%\begin{IEEEproof}
%If $I(X_2; Y_2) \leq I(X_2; Y_1)$ holds for the pmf $Q_2^* = \arg \max\limits_{\{p(x_2)\}} I(X_2; Y_2)$, then from the lower bound~(\ref{eqnDefLBblock}) we have, for any pmf $p(x_2)$, min
%\end{IEEEproof}

{\bf Remarks.}
\begin{enumerate}
\item As seen from~(\ref{eqnDefUppBound}) and~(\ref{eqnDefLowBound}), the initial state of the GTC is involved in the derived bounds. This cannot be avoided if the initial state is not known to the secondary users or if the access time\footnote{By a time we mean a stage of the trellis that represents the GTC.} of the secondary users is at a random time $t$ instead of $t= 0$. However, most GTCs~(for example, block codes and convolutional codes) satisfy that the process $\{S_t\}$ takes the uniform distribution as the unique stationary distribution. Equivalently, a sufficiently long i.u.d. sequence of inputs at User A can drive the GTC encoder from any given initial state into each state with equal probability. In this case, both the links ${\bf X}_2\rightarrow {\bf Y}_2$ and ${\bf X}_2\rightarrow {\bf Y}_1$ are {\em indecomposable}~\cite{Gallager68}. Then the initial state can be fixed in the derived bounds.
\item The readers may notice that the error performance requirement $\varepsilon$ by the primary users is not shown explicitly in the above derived bounds.
    However, these bounds do depend on the parameter $\varepsilon$ because the involved mutual information rates depend on the primary transmission power $P_1$ that is closely related to the primary error performance requirement $\varepsilon$. In particular, in the case of $\delta_1 = 1$, the primary transmission power $P_1$ plays the same role as the error performance requirement $\varepsilon$ does. In a general case, we may use~(say) $C_L(\varepsilon, \delta_1, P_2)$ to emphasize the dependence of the lower bound on $\varepsilon$, $\delta_1$ and $P_2$.
\end{enumerate}
%since both links ${\bf X}_2 \rightarrow {\bf Y}_2$ and ${\bf X}_2 \rightarrow {\bf Y}_1$ are affected by the transmission power $P_1$, while $P_1$ is closely related to $\varepsilon$.
%

%\subsection{Lower Bounds for the Accessible Capacity in the case of Non-trivial Interference Margin}
%\subsection{Non-trivial Interference Margin (i.e., $\delta_1 > 1$)}\label{SubsecNonTriIM}
\subsection{Non-trivial Interference Margin}\label{SubsecNonTriIM}

%To get insight into the effect of the error performance requirement by the primary users $\varepsilon$ on the accessible capacity, we assume that the interference margin at the primary receiver is non-trivial and cannot be removed,

Now let us discuss the case when non-trivial interference margin exists at the primary transmitter, i.e., $\delta_1 > 1$.
Let $\Delta P_2 = \min\{P_2, (\delta_1 - 1) / a_{21}^2\}$ and $P'_2 = P_2 - \Delta P_2$. Define ${\bf X}_2 = {\bf X'}_2 + {\bf \Delta X}_2$, where ${\bf X'}_2$ is a discrete random sequence over $\mathcal{X}_2^N$ with pmf $p({\bf x'}_2)$ and power constraint $P'_2$, i.e., ${\bf E}\left[\|{\bf X'}_2\|^2\right] \leq nN P'_2$, and ${\bf \Delta X}_2$ is a Gaussian random sequence with mean zero and variance $\Delta P_2$ per dimension. Then the considered system in~(\ref{eqnIFCseqRV}) can be expressed as
\begin{equation}\label{eqnIFCseqRV1}
   \begin{array}{ccrcrcrcl}
    {\bf Y}_1 & = & {\bf X}_1 & + & a_{21} {\bf X'}_2 & + & a_{21} {\bf \Delta X}_2 & + & {\bf Z}_1\\
    {\bf Y}_2 & = & a_{12} {\bf X}_1 & + & {\bf X'}_2 & + & {\bf \Delta X}_2 & + & {\bf Z}_2
   \end{array}.
\end{equation}

%\begin{theorem}\label{THCapLBnon-tri}
%Let $R_2$ be the rate $R_2 = R_{21} + R_{22}$ satisfying
%\begin{equation}\label{eqnDefLBnon-tri1}
%R_{21} \leq C_L(\varepsilon, 1, P'_2) \stackrel{\Delta}{=} \lim_{N\rightarrow \infty} \frac{1}{nN} \max_{\{p({\bf x'}_2)\}}\min_{s_0}
%    \min\left\{I({\bf X'}_2; {\bf Y}_1|s_0), I({\bf X'}_2; {\bf Y}_2|s_0)\right\}
%\end{equation}
%and
%\begin{equation}\label{eqnDefLBnon-tri2}
%%R_{22} \leq \frac{1}{nN} \left(h(a_{12}{\bf X}_1 + {\bf \Delta X}_2 + {\bf Z}_2) - h(a_{12}{\bf X}_1 + {\bf Z}_2)\right).
%R_{22} \leq \frac{1}{nN} I\!\left(\!\left.{\bf \Delta X}_2;{\bf Y}_2\right|{\bf X'}_2\right).
%\end{equation}
%Then $R_2$ is accessible.
%\hfill \ding{113}
%\end{theorem}

\begin{theorem}\label{THCapLBnon-tri}
Let $R_2 = R_{21} + R_{22}$ satisfying
\begin{equation}\label{eqnDefLBnon-tri1}
R_{21} \leq C_L(\varepsilon, 1, P'_2) \stackrel{\Delta}{=} \lim_{N\rightarrow \infty} \frac{1}{nN} \max_{\{p({\bf x'}_2)\}}\min_{s_0}
    \min\left\{I({\bf X'}_2; {\bf Y}_1|s_0), I({\bf X'}_2; {\bf Y}_2|s_0)\right\}
\end{equation}
and
\begin{equation}\label{eqnDefLBnon-tri2}
%R_{22} \leq \frac{1}{nN} \left(h(a_{12}{\bf X}_1 + {\bf \Delta X}_2 + {\bf Z}_2) - h(a_{12}{\bf X}_1 + {\bf Z}_2)\right).
R_{22} \leq \lim_{N\rightarrow \infty} \frac{1}{nN} I\!\left(\!\left.{\bf \Delta X}_2;{\bf Y}_2\right|{\bf X'}_2\right).
\end{equation}
Then $R_2$ is accessible.
\hfill \ding{113}
\end{theorem}

\begin{IEEEproof}%The primary user employs the fixed GTC $\mathscr{C}_1$ as described in Sec.~\ref{SecProb}.
Here we follow Han and Kobayashi's idea of splitting message into the common and private parts~\cite{Han81}. That is, the message set at the secondary transmitter is assumed to be $\mathcal{M}_2 = \mathcal{L}_1 \times \mathcal{L}_2$, where $\mathcal{L}_1 = \{1,2, \ldots, L_1\}$ is the common message and will be decoded at both receivers and $\mathcal{L}_2 = \{1,2, \ldots, L_2\}$ is the private message set and will be decoded only at the secondary receiver. Obviously, $M_2 = L_1 L_2$.

{\em \underline{Codebook Generation}:} Generate $L_1$ independent codewords ${\bf x'}_2 \in \mathcal{X}_2^{N}$ at random according to the pmf $p({\bf x'}_2)$. Generate $L_2$ independent codewords $\Delta {\bf x}_2 \in \mathbb{R}^{nN}$ at random by drawing each codeword's coordinates independently according to the Gaussian distribution $\mathcal{N}(0, \Delta P_2)$.
For sufficiently large $N$, without loss of optimality, we can assume that these two codebooks satisfy power constraints $\mathbf{E}\left[\|{\bf X'}_2\|^2\right] \leq nNP'_2$ and $\mathbf{E}\left[\|{\bf \Delta X}_2\|^2\right] \leq nN\cdot \Delta P_2$, respectively.

{\em \underline{Encoding}:} The encoding function at User C is
\begin{eqnarray}
    \begin{array}{cccl}
        \phi_2: & \mathcal{M}_2 & \rightarrow & \mathbb{R}^{nN} \\
           & w_2=(v_1, v_2)  &\mapsto & \phi_2(w_2) = {\bf x}_2(w_2) = {\bf x'}_2(v_1) + {\bf \Delta x}_2(v_2)
    \end{array}
\end{eqnarray}
where $v_1 \in \mathcal{L}_1$ and $v_2 \in \mathcal{L}_2$ are the common message and the private message transmitted by the secondary user, respectively.

%{\em \underline{Decoding}:} After transmission over the system~(\ref{eqnIFCseqRV1}), the received sequences at User~B and~D can be rewritten as
%\begin{equation}\label{eqnIFCseqRV2}
%           \begin{array}{ccrcrcl}
%            {\bf y}_1 & = & {\bf x}_1 & + & a_{21} {\bf x'}_2 & + & {\bf z'}_1\\
%            {\bf y}_2 & = & a_{12} {\bf x}_1 & + & {\bf x'}_2 & + & {\bf z'}_2
%           \end{array}
%\end{equation}
%where ${\bf z'}_1 = a_{21} {\bf \Delta x}_2 + {\bf z}_1$ and ${\bf z'}_2 = {\bf \Delta x}_2 + {\bf z}_2$. Treating ${\bf z'}_1$ and ${\bf z'}_2$ as noises, we find that the above system~(\ref{eqnIFCseqRV2}) has trivial interference margin when $P_2 > (\delta_1 - 1) / a_{21}^2$~(which implies $P'_2 > 0$)~\footnote{If $P_2 \leq (\delta_1 - 1) / a_{21}^2$, then $P'_2 = 0$ and $C_L(\varepsilon,1,P'_2)=0$.}. Applying the result in the previous subsection directly to the system~(\ref{eqnIFCseqRV2}), we know that any rate $R_{21} \stackrel{\Delta}{=} \frac{\log L_1}{nN}$ such that $R_{21} \leq C_L(\varepsilon,1,P'_2)$~(defined in~(\ref{eqnDefLBnon-tri1})) is accessible.
%From the previous subsection, we know that the secondary common message $v_1$ transmitted by the codeword ${\bf x'}_2(v_1)$ can be decoded at both receivers, and the primary message $w_1$ can be decoded by using a (partial) interference cancellation successive decoding.

{\em \underline{Decoding}:} After transmission over the system~(\ref{eqnIFCseqRV1}), the received sequences at User~B and~D can be rewritten as
\begin{equation}\label{eqnIFCseqRV2}
           \begin{array}{ccrcrcl}
            {\bf y}_1 & = & {\bf x}_1 & + & a_{21} {\bf x'}_2 & + & {\bf z'}_1\\
            {\bf y}_2 & = & a_{12} {\bf x}_1 & + & {\bf x'}_2 & + & {\bf z'}_2
           \end{array}
\end{equation}
where ${\bf z'}_1 = a_{21} {\bf \Delta x}_2 + {\bf z}_1$ and ${\bf z'}_2 = {\bf \Delta x}_2 + {\bf z}_2$. Since $R_{21} \stackrel{\Delta}{=} \frac{\log L_1}{nN} \leq C_L(\varepsilon,1,P'_2)$~(defined in~(\ref{eqnDefLBnon-tri1})), we can prove~(following the proof of Theorem~\ref{THCapBound}) by treating ${\bf z'}_1$ and ${\bf z'}_2$ as noises that the secondary common message $v_1$ transmitted by the codeword ${\bf x'}_2(v_1)$ can be decoded correctly with high probability at both receivers.
After subtracting the decoded codeword $a_{21}{\bf x'}_2(v_1)$ from ${\bf y}_1$, we can verify that the SNR at the primary receiver is $P_1/(1 + a_{21}^2 \Delta P_2) \geq P_1^*$ and hence the primary decoding performance will not get worse.
After subtracting the decoded codeword ${\bf x'}_2(v_1)$ from ${\bf y}_2$, the secondary receiver can decode the private message $v_2$ correctly with high probability since $R_{22} \leq \lim_{N\rightarrow \infty} \frac{1}{nN} I\!\left(\!\left.{\bf \Delta X}_2;{\bf Y}_2\right|{\bf X'}_2\right)$.
\end{IEEEproof}

%~(equivalently, $R_{22} \leq \lim_{N\rightarrow \infty} \frac{1}{nN} \left(h(a_{12}{\bf X}_1 + {\bf \Delta X}_2 + {\bf Z}_2) - h(a_{12}{\bf X}_1 + {\bf Z}_2)\right)$).

The proof of Theorem~\ref{THCapLBnon-tri} is different from that of Theorem~\ref{THCapBound} where no inference margin exists: 1)~the secondary transmitter uses a two-level superposition~(rather than single-level) code; 2)~the primary decoder uses a two-stage successive {\em partial}~(rather than {\em full}) interference cancellation decoding algorithm, while the secondary decoder uses a two-stage successive~(rather than one-stage) decoding algorithm.

Theorem~\ref{THCapLBnon-tri} indicates that, when non-trivial interference margin exists at the primary receiver, the accessible capacity can be lower-bounded by
\begin{equation}\label{eqnDefLBnon-tri3}
C_L(\varepsilon,\delta_1,P_2) \stackrel{\Delta}{=} C_L(\varepsilon, 1, P'_2) +
\lim_{N\rightarrow \infty} \frac{1}{nN} I\!\left(\!\left.{\bf \Delta X}_2;{\bf Y}_2\right|{\bf X'}_2\right).
\end{equation}

{\bf Remark.}
Given $\mathscr{C}_1$ and $P_1$, let $\varepsilon_{\min}$ be the error performance requirement corresponding to the interference margin $\delta_1 = 1$. That is, $\varepsilon_{\min}$ can not be decreased further unless $P_1$ is increased or the GTC $\mathscr{C}_1$ is replaced by a better one. Intuitively, as the error performance requirement is relaxed~(i.e., $\varepsilon$ increases from $\varepsilon_{\min}$), the interference margin is getting larger~(i.e., $\delta_1$ increases). This means that the primary users can tolerate stronger interference. As a result, the accessible capacity of the secondary users can not decrease as the error performance requirement by the primary users is relaxed. However, the accessible capacity must be upper bounded by the maximum secondary transmission rate in the case when the error performance requirement by the primary users is not considered at all\footnote{This can happen when the secondary users are illegal and ``rude''.}.

\section{The Evaluation of the Upper and Lower Bounds}\label{SecEvaBounds}

We have now derived upper and lower bounds on the accessible capacity for the considered GIFC~(\ref{eqnIFCseqRV}), as shown in Theorems~\ref{THCapBound} and~\ref{THCapLBnon-tri}. However, it could be very complicated to evaluate these bounds. The difficulty arises from the following two facts. Firstly, a complex GTC for primary users that has a large number of branches must result in intractable FSCs for the secondary links. Secondly, even if the GTC is simple, the space of sequences ${\bf x}_2$ expands exponentially as the length $N$ increases, which makes it infeasible to optimize the mutual information rates over all possible pmfs $p({\bf x}_2)$ for large $N$. Therefore, we consider only those tractable GTCs and assume that the transmitted signal sequences ${\bf X}_2 \in \mathcal{X}_2^N$ are i.u.d. under the power constraint $P_2$. The resulting accessible rate is referred to as i.u.d. accessible capacity and denoted by $C_a^{(i.u.d.)}$.
The i.u.d. accessible capacity plays an important role in the practice of coding design since it specifies the achievable rate when random linear~(coset) codes are implemented, as shown in~\cite[Theorems~1 and~2]{Kavcic05}.
The objective of this section is to evaluate lower bounds and upper bounds on the i.u.d. accessible capacity.
In what follows, we have fixed the initial state as $s_0 = 0$ and removed it from the equations for simplicity.

\subsection{Computations of the Mutual Information Rates}\label{SubSecComRats}

From Theorems~\ref{THCapBound} and~\ref{THCapLBnon-tri}, we have bounds as follows.
\begin{enumerate}
    \item In the case when no interference margin exists at the primary receiver, the i.u.d. accessible capacity $C_a^{(i.u.d.)}$ can be upper-bounded and lower-bounded by
        \begin{equation}\label{eqnUB-IUD}
            C_U^{(i.u.d.)} = \lim_{N \rightarrow \infty}\frac{1}{nN} I({\bf X}_2; {\bf Y}_2)
        \end{equation}
        and
        \begin{equation}\label{eqnLB-IUD}
            C_L^{(i.u.d.)} = \lim_{N \rightarrow \infty}\frac{1}{nN} \min\{I({\bf X}_2; {\bf Y}_1), I({\bf X}_2; {\bf Y}_2)\},
        \end{equation}
        respectively, where the channel input sequence ${\bf X}_2$ is i.u.d. according to the pmf $p({\bf x}_2) = \frac{1}{|\mathcal{X}_2|^N}$. Obviously, $C_U^{(i.u.d.)}$ is a lower bound of the derived upper bound $C_U$ defined in Lemma~\ref{LMlimits} and hence it may not be an upper bound on the accessible capacity $C_a$, while $C_L^{(i.u.d.)}$ does serve as a lower bound on $C_a$.

    \item In the case when non-trivial interference margin exists at the primary receiver, the accessible capacity $C_a$ can be lower-bounded by
        \begin{equation}\label{eqnLBnon-trivIUD}
        {C'}_L^{(i.u.d.)} = \lim\limits_{N\rightarrow \infty} \frac{1}{nN}
            \left(I\!\left(\!\left.{\bf \Delta X}_2;{\bf Y}_2\right|{\bf X'}_2\right)
            + \min \left\{I({\bf X'}_2; {\bf Y}_1), I({\bf X'}_2; {\bf Y}_2)\right\}\right)
        \end{equation}
        where the channel input sequence ${\bf X}_2$ is a superposition of an i.u.d. sequence ${\bf X'}_2$ according to the pmf $p({\bf x'}_2) = \frac{1}{|\mathcal{X}_2|^N}$ and an i.i.d. sequence ${\bf \Delta X}_2$ according to the Gaussian distribution $\mathcal{N}(0,\Delta P_2)$.
\end{enumerate}

The above upper and lower bounds can be evaluated by adapting the methods used in~\cite{Arnold01,Pfister01,Sharma01,Arnold06}.
We focus on the evaluation of the upper bound, while the lower bounds can be estimated similarly. Specifically, we can express the upper bound $C_U^{(i.u.d.)}$ as
\begin{eqnarray}
    C_U^{(i.u.d.)}
    &=&  \lim_{N \rightarrow \infty} \frac{1}{nN} I({\bf X}_2; {\bf Y}_2)\nonumber\\
    &=&  \lim_{N \rightarrow \infty} \frac{1}{nN} h({\bf Y}_2) - \lim_{N \rightarrow \infty} \frac{1}{nN} h({\bf Y}_2|{\bf X}_2)\nonumber\\
    &=&  \lim_{N \rightarrow \infty} \frac{1}{nN} h({\bf Y}_2) - \lim_{N \rightarrow \infty}\frac{1}{nN} h(a_{12}{\bf X}_1 + {\bf Z}_2).
    %&=& \frac{1}{nN} \sum_{t=1}^N h(Y_{2,t} | Y_{2}^{(t-1)}) - \frac{1}{nN} h(Y_{2,t} | Y_{2}^{(t-1)}, {\bf X}_2)\\
\end{eqnarray}
These two entropy rates $\lim\limits_{N \rightarrow \infty} \frac{1}{nN} h({\bf Y}_2)$ and $\lim\limits_{N \rightarrow \infty} \frac{1}{nN} h(a_{12}{\bf X}_1 + {\bf Z}_2)$ can be computed by a similar method since both ${\bf Y}_2$ and $a_{12}{\bf X}_1 + {\bf Z}_2$ can be viewed as hidden Markov chains. As an example, we show how to compute $\lim\limits_{N \rightarrow \infty} \frac{1}{nN} h({\bf Y}_2)$ in the following.

\begin{figure}
  \centering
  \includegraphics[width=12.5cm]{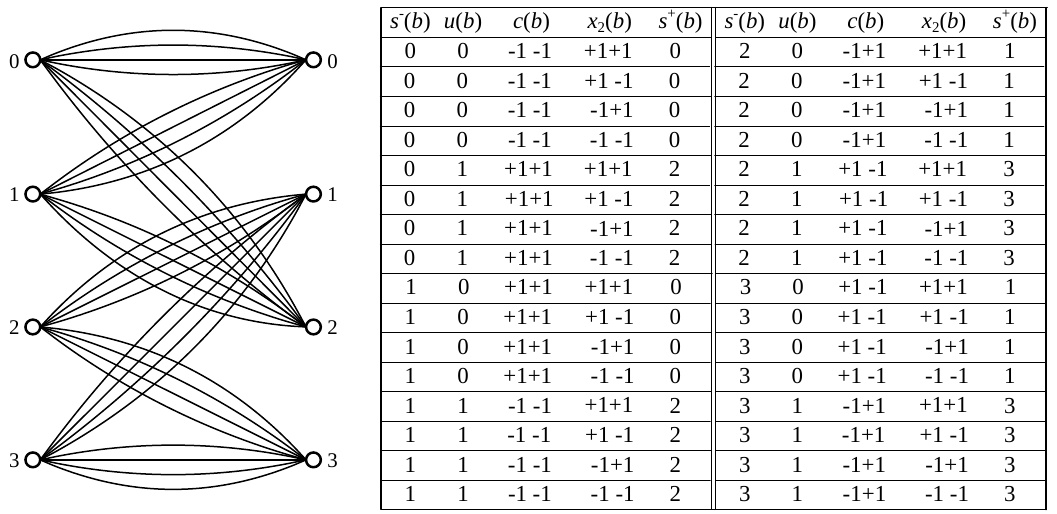}\\
  \caption{A trellis section of the link ${\bf X}_2 \rightarrow {\bf Y}_2$ where the GTC $\mathscr{C}_1$ represents the (2,1,2) convolutional coded BPSK shown in Example~\ref{ExCCBPSK}.}\label{FigTRIFFC}
\end{figure}

From the Shannon-McMillan-Breiman theorem~\cite[Theorem~15.7.1]{Cover91}, we know that, with probability $1$,
\begin{equation}
    \lim_{N \rightarrow \infty} -\frac{1}{nN} \log f({\bf y}_2) = \lim_{N \rightarrow \infty} \frac{1}{nN} h({\bf Y}_2),
\end{equation}
since the sequence ${\bf Y}_2$ is a stationary stochastic process. Then evaluating $ \lim\limits_{N \rightarrow \infty} \frac{1}{nN} h({\bf Y}_2)$ is converted to computing
%$\lim\limits_{N \rightarrow \infty} -\frac{1}{nN} \log f({\bf y}_2)$ as
\begin{equation}\label{eqnLogSum}
    \lim\limits_{N \rightarrow \infty} -\frac{1}{nN} \log f({\bf y}_2)
    \approx -\frac{1}{nN} \log \left(\prod_{t=1}^N f(y_{2,t}|y_2^{(t-1)}) \right)
    = -\frac{1}{nN} \sum_{t=1}^N \log f(y_{2,t}|y_2^{(t-1)})
\end{equation}
for a sufficiently long {\em typical} sequence ${\bf y}_2$. Then, the key is to compute the conditional probabilities $f(y_{2,t}|y_2^{(t-1)})$ for all $t$. This can be done by performing the BCJR algorithm over a new~(time-invariant) trellis, which is constructed by modifying the original trellis of the GTC $\mathscr{C}_1$. Actually, the link ${\bf X}_2 \rightarrow {\bf Y}_2$ can be represented by the following modified trellis.
\begin{itemize}
    \item The new trellis has the same state set $\mathcal{S} = \{0,1,\ldots,|\mathcal{S}| -1\}$ as the GTC $\mathscr{C}_1$.
    \item Each branch $b=(s^{-}(b), u(b), c(b), s^{+}(b))$ in $\mathcal{B}$ of the original trellis is expanded into $|\mathcal{X}_2|$ parallel branches
        \begin{equation}
            \left\{\!\left.b \!=\! (s^{-}(b), u(b), c(b), x_2(b), s^{+}(b))\right|x_2(b) \!\in\! \mathcal{X}_2 {\rm~is~the~transmitted~signal~at~User~C}\right\}.
        \end{equation}
        %The new branch set, denoted by $\bar{\mathcal{B}}$, has size of $|\mathcal{B}| \times |\mathcal{X}_2|$.
\end{itemize}
Fig.~\ref{FigTRIFFC} depicts a trellis section of the link ${\bf X}_2 \rightarrow {\bf Y}_2$, where the GTC $\mathscr{C}_1$ is the $(2,1,2)$-CCBPSK as introduced in Example~\ref{ExCCBPSK}.

%------------------------------------------------------------------------------------------------------
%------------------------------------------------------------------------------------------------------

Given the received sequence ${\bf y}_2$ at User D, similar to the BCJR algorithm~\cite{BCJR74}, we define

{\em Branch Metrics:} To each branch $b_t = (s_{t-1}, u_t, c_t, x_{2,t}, s_{t})$, we assign a metric
\begin{eqnarray}
    \rho(b_t)
    &\stackrel{\Delta}{=}& f(b_t, y_{2,t}| s_{t-1})\nonumber\\
    &=& p(u_t)\, p(x_{2,t})\, f(y_{2,t} | s_{t-1}, u_t, x_{2,t})\nonumber\\
    &=& \frac{1}{M} \frac{1}{|\mathcal{X}_2|} \frac{1}{{(2\pi)}^{n/2}} \exp\left\{-\frac{\| y_{2,t} - a_{12} \sqrt{P_1} c_t- x_{2,t} \|^2}{2}\right\}.
\end{eqnarray}

{\em State Transition Probabilities:} The transition probability from $s_{t-1}$ to $s_t$ is defined as
\begin{eqnarray}
    \gamma_t(s_{t-1}, s_{t})
    &\stackrel{\Delta}{=}& f\left(s_t, y_{2,t} | s_{t-1}\right)\\
    &=& \sum_{b:s^{-}(b) = s_{t-1}, s^{+}(b) = s_t} \rho(b).
\end{eqnarray}

{\em Forward Recursion Variables:}
We define the posterior probabilities
\begin{equation}
    \alpha_t(s_t) \stackrel{\Delta}{=} p\left(s_t | y_2^{(t)} \right), \;\;t=0,1,\ldots, N.
\end{equation}
Then
\begin{equation}\label{eqnProbYcond}
    f(y_{2,t} | y_2^{(t-1)}) = \sum_{s_{t-1},s_t} \alpha_{t-1}(s_{t-1}) \gamma_t(s_{t-1}, s_{t}),
\end{equation}
where the values of $\alpha_t(s_t)$ can be computed recursively by%by the forward recursion of the BCJR algorithm as follows
\begin{equation}\label{eqnAlpha}
    \alpha_t(s_t)
    = \frac{\sum_{s_{t-1}} \alpha_{t-1}(s_{t-1}) \gamma_t(s_{t-1}, s_{t})}
        {\sum_{s_{t-1},s_t} \alpha_{t-1}(s_{t-1}) \gamma_t(s_{t-1}, s_{t})}.
\end{equation}
In summary, the algorithm to estimate the entropy rate $\lim\limits_{N\rightarrow \infty} \frac{1}{nN}h({\bf Y}_2)$ is described as follows.
\begin{algorithm}\label{Alg}
\hspace{\parindent}
\begin{enumerate}
    \item {\bf Initializations:} Choose a sufficiently large number $N$. Set the initial state of the GTC to be $s_0 = 0$. The forward recursion variables are initialized as $\alpha_0(s) = 1$ if $s=0$ and otherwise $\alpha_0(s) = 0$.

    \item {\bf Simulations for User A:}
        \begin{enumerate}
            \item Generate a message sequence $(u_1,u_2,\ldots,u_N)$ independently according to the uniform distribution $p(U = u_i) = \frac{1}{M}$.
            \item Encode the message sequence by the encoder of the GTC $\mathscr{C}_1$ and get the coded sequence $(c_1,c_2,\ldots,c_N)$.

            \item Transmit the signal sequence ${\bf x}_1 = \sqrt{P_1}(c_1,c_2, \ldots, c_N)$.
        \end{enumerate}
    \item {\bf Simulations for User C:}
        \begin{enumerate}
            \item Generate a sequence ${\bf x}_2 \in \mathcal{X}_2^{N}$ independently according to the uniform distribution $p(x_2) = \frac{1}{|\mathcal{X}_2|}$.
            \item Transmit the signal sequence ${\bf x}_2$.
        \end{enumerate}

    \item {\bf Simulations for User D:}
        \begin{enumerate}
            \item Generate an $(nN)$-sequence ${\bf z}_2 \in \mathbb{R}^{nN}$ independently according to the Gaussian distribution $\mathcal{N}(0,1)$.

            \item Receive the sequence ${\bf y}_2 = a_{12}{\bf x}_1 + {\bf x}_2 + {\bf z}_2$.
        \end{enumerate}
    \item {\bf Computations:}
        \begin{enumerate}
            \item For $t=1,\ldots, N$, compute the values of $f(y_{2,t}|y_2^{(t-1)})$ and $\alpha_t(s_t)$ recursively according to equations~(\ref{eqnProbYcond}) and~(\ref{eqnAlpha}).

            \item Evaluate the entropy rate
                \begin{equation}
                    \lim_{N \rightarrow \infty} \frac{1}{nN} h({\bf Y}_2) \approx - \frac{1}{nN} \sum_{t=1}^N \log(f(y_{2,t}|y_2^{(t-1)})).
                \end{equation}
        \end{enumerate}
\end{enumerate}
\hfill \ding{113}
\end{algorithm}

Similarly, we can evaluate the entropy rate $\lim\limits_{N \rightarrow \infty} \frac{1}{nN} h(a_{12}{\bf X}_1 + {\bf Z}_2)$. Therefore, we obtain the bound as
\begin{equation}
    C_U^{(i.u.d.)} = \lim_{N \rightarrow \infty} \frac{1}{nN} h({\bf Y}_2) - \lim_{N \rightarrow \infty} \frac{1}{nN} h(a_{12}{\bf X}_1 + {\bf Z}_2).
\end{equation}

\subsection{Numerical Results and Discussion}\label{SubSecNumRes}

In this subsection, we give numerical results to show the dependence of the accessible capacity on parameters.
In all simulations, we assume that both User A and C utilize BPSK modulation since we are primarily concerned with the low-SNR regime, and we choose a large number $N=10^6$.

\subsubsection{The Impact of the GTCs on Accessible Rates}
To investigate the dependence of the accessible rates on GTCs, we assume that no interference margin exists at the primary receiver.
The power at User A is fixed to be $P_1 = 6$, while the power $P_2$ at User C is allowed to be varied. The GTCs at User A we simulated are listed in Table~\ref{TablComprCodes}, where the first four GTCs are the examples introduced in Sec.~\ref{SubSecPL} and the last GTC is the $(3,1,2)$ convolutional code~(defined by the generator matrix $G(D)=[1+D \;\; 1+D^2 \;\; 1+D+D^2]$) with the BPSK signaling.
Also given in Table~\ref{TablComprCodes} are the error performances corresponding to the fixed transmission power $P_1 = 6$.
These error performances will be getting worse if treating the interference from the secondary users as noise. In this case,
the upper and lower bounds $C_U^{(i.u.d.)}$ and $C_L^{(i.u.d.)}$, given in~(\ref{eqnUB-IUD}) and~(\ref{eqnLB-IUD}), are computed.

The GIFCs we simulated have interference coefficients $a_{12}^2 = a_{21}^2 = 1.5$~(strong interference), $a_{12}^2 = a_{21}^2 = 1$~(median interference), $a_{12}^2 = a_{21}^2 = 0.5$~(weak interference), or $a_{12}^2 = 0.5, a_{21}^2 = 1.5$~(asymmetric interference).

\begin{table}
  \centering
  \caption{Comparison of the GTCs}\label{TablComprCodes}
  \begin{tabular}{|l|c|c|c|}
    \hline
    GTC & Rate & $P_1$ & $\varepsilon$ \\ \hline
    Uncoded-BPSK & 1 & 6 & $\approx 0.72\,e\!-\!2$ \\
    $[2,1,2]$-RCBPSK & 1/2 & 6 & $\approx 0.28\,e\!-\!3$ \\
    $[8,4,4]$-EHCBPSK & 1/2 & 6 & $\approx 0.20\,e\!-\!4$ \\
    $(2,1,2)$-CCBPSK & 1/2 & 6 & $\approx 0.63\,e\!-\!7$ \\
    $(3,1,2)$-CCBPSK & 1/3 & 6 & $ < 0.63\,e\!-\!7$ \\
    \hline
  \end{tabular}
\end{table}
%
%\begin{table}
%  \centering
%  \caption{Comparison of the GTCs}\label{ComprCodes1}
%  \begin{tabular}{|l|c|c|c|}
%    \hline
%    % after \\: \hline or \cline{col1-col2} \cline{col3-col4} ...
%    Code & Rate & $P_1$ & BER \\ \hline
%    Uncoded-BPSK & 1 & 2 & $\approx 0.78\,e\!-\!1$ \\
%    $[2,1,2]$-RCBPSK & 1/2 & 2 & $\approx 0.23\,e\!-\!1$ \\
%    $[8,4,4]$-EHCBPSK & 1/2 & 2 & $\approx 0.15\,e\!-\!1$ \\
%    $(2,1,2)$-CCBPSK & 1/2 & 2 & $\approx 0.34\,e\!-\!2$ \\
%    $(3,1,2)$-CCBPSK & 1/3 & 2 & $\approx 0.3\,e\!-\!4$ \\
%    \hline
%  \end{tabular}
%\end{table}
%
%\begin{table}
%  \centering
%  \caption{Comparison of the GTCs}\label{ComprCodes2}
%  \begin{tabular}{|l|c|c|c|}
%    \hline
%    % after \\: \hline or \cline{col1-col2} \cline{col3-col4} ...
%    Code & Rate & $P_1$ & BER \\ \hline
%    Uncoded-BPSK & 1 & 0.5 & $\approx 0.24$ \\
%    $[2,1,2]$-RCBPSK & 1/2 & 0.5 & $\approx 0.16$ \\
%    $[8,4,4]$-EHCBPSK & 1/2 & 0.5 & $\approx 0.21$ \\
%    $(2,1,2)$-CCBPSK & 1/2 & 0.5 & $\approx 0.3$ \\
%    $(3,1,2)$-CCBPSK & 1/3 & 0.5 & $\approx 0.17$ \\
%    \hline
%  \end{tabular}
%\end{table}

\begin{figure}[!t]
  \centering
  \includegraphics[width=10.0cm]{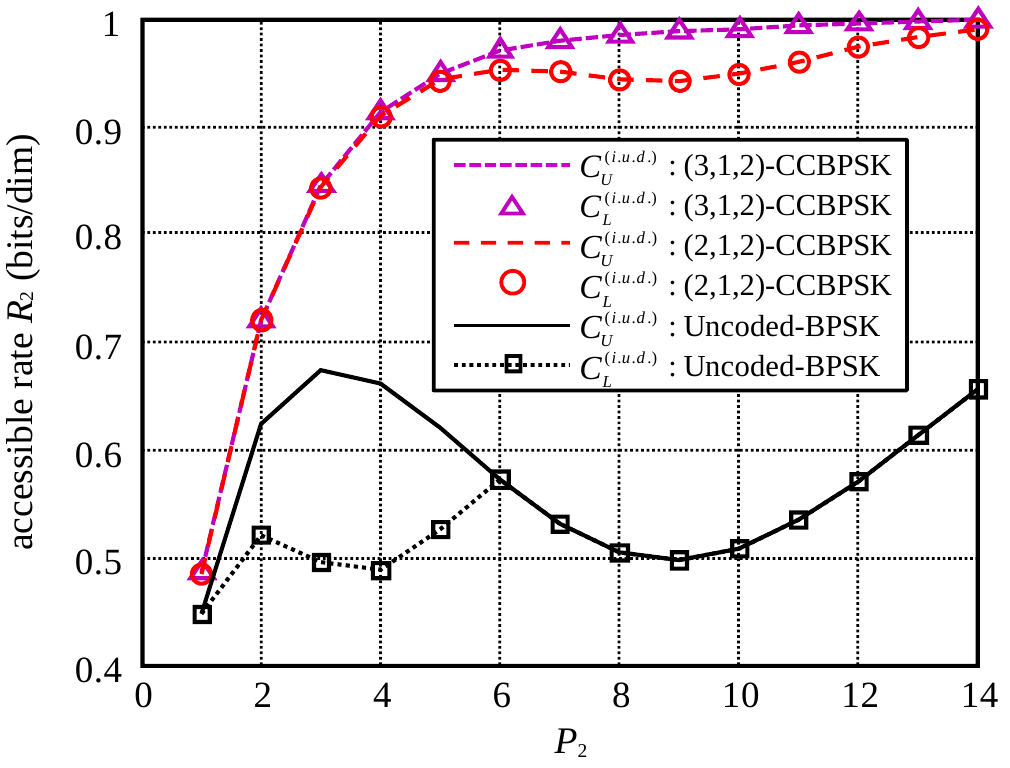}\\
  \caption{Bounds on the i.u.d. accessible capacities of the GIFC with $a_{12}^2 = a_{21}^2 = 1.5$ for different GTCs with different coding rates.}\label{FigUnCoStrong}
\end{figure}

\begin{figure}[!t]
  \centering
  \includegraphics[width=10.0cm]{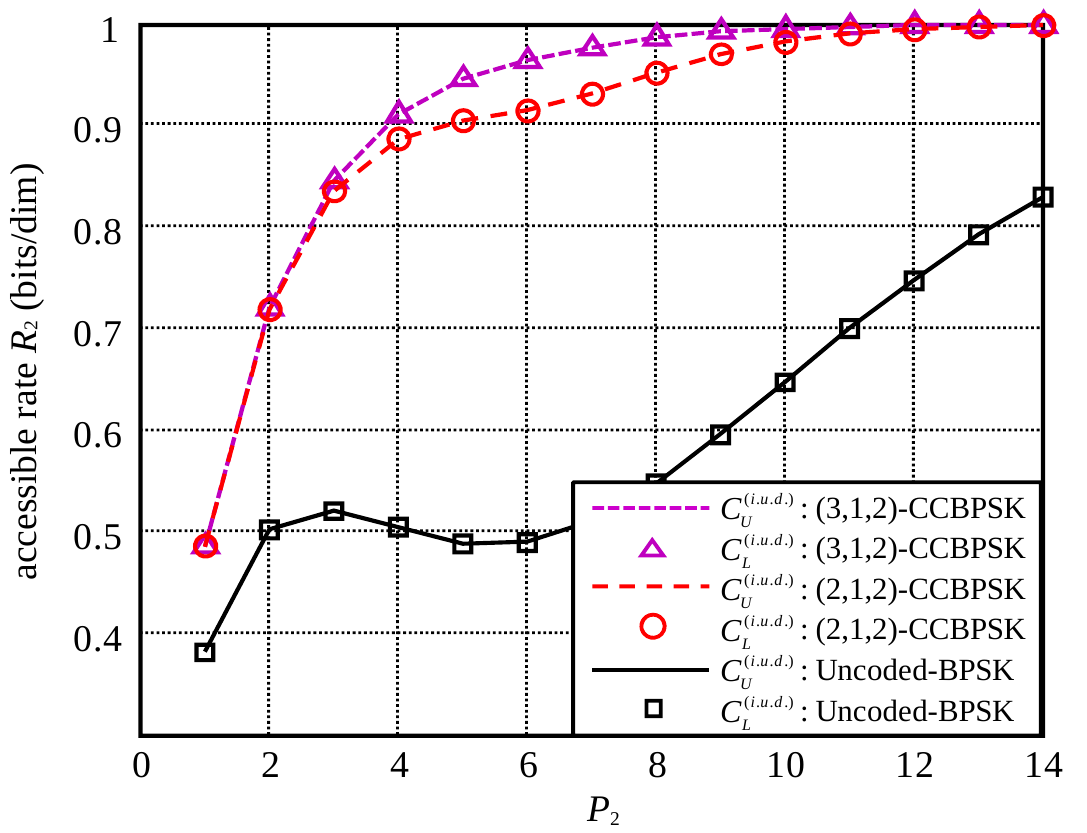}\\
  \caption{Bounds on the i.u.d. accessible capacities of the GIFC with $a_{12}^2 = a_{21}^2 = 1$ for different GTCs with different coding rates.}\label{FigUnCoMid}
\end{figure}

\begin{figure}[!t]
  \centering
  \includegraphics[width=10.0cm]{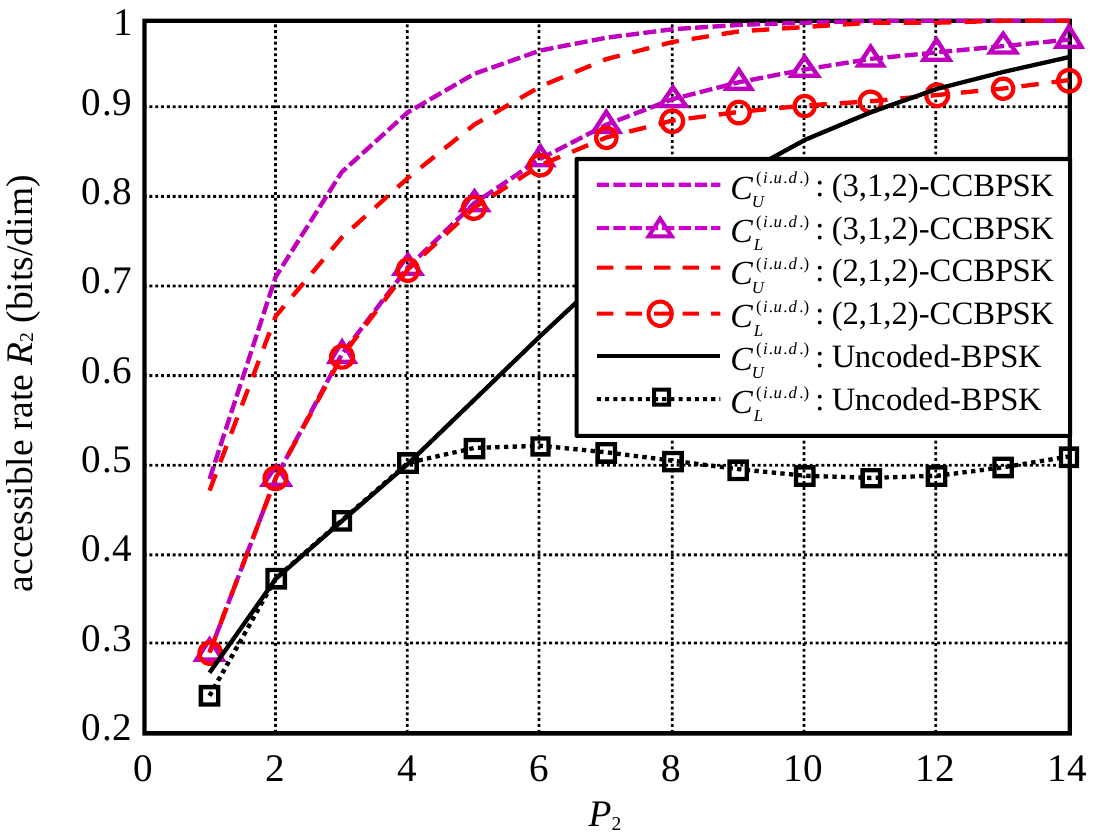}\\
  \caption{Bounds on the i.u.d. accessible capacities of the GIFC with $a_{12}^2 = a_{21}^2 = 0.5$ for different GTCs with different coding rates.}\label{FigUnCoWeak}
\end{figure}

\begin{figure}[!t]
  \centering
  \includegraphics[width=10.0cm]{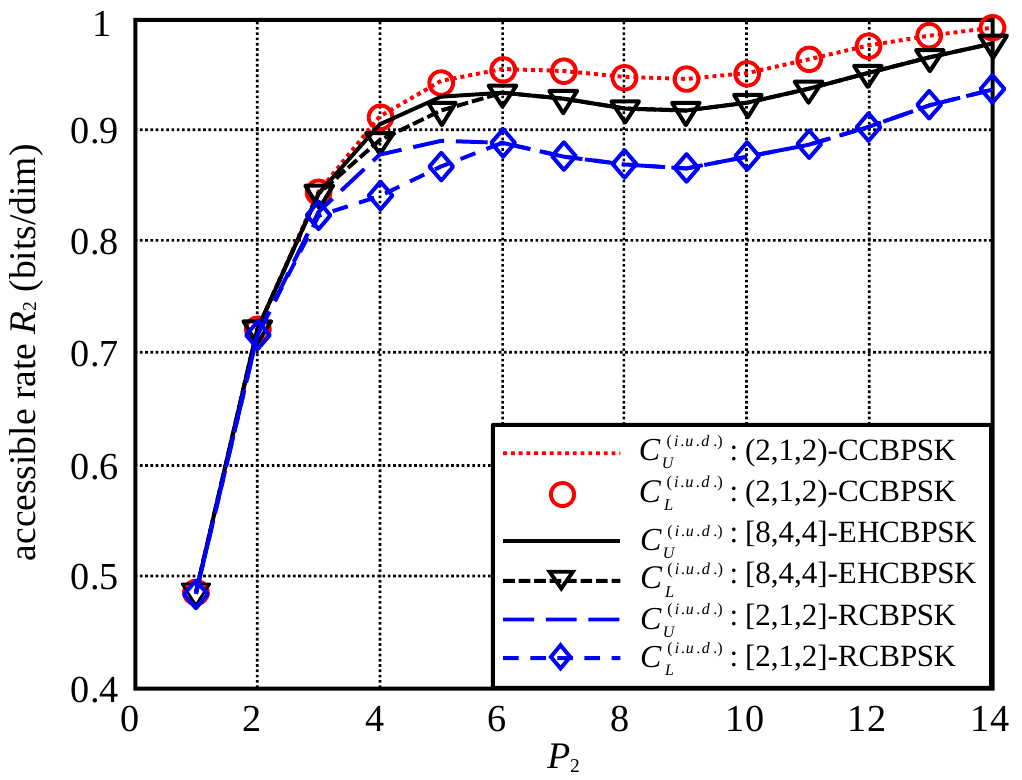}\\
%  \caption{The interference coefficients of the considered GIFC are $a_{12}^2 = a_{21}^2 = 1.5$.}\label{FigHalfRStrong}
  \caption{Bounds on the i.u.d. accessible capacities of the GIFC with $a_{12}^2 = a_{21}^2 = 1.5$ for different GTCs with the same coding rate $1/2$.}\label{FigHalfRStrong}
\end{figure}

\begin{figure}[!t]
  \centering
  \includegraphics[width=10.0cm]{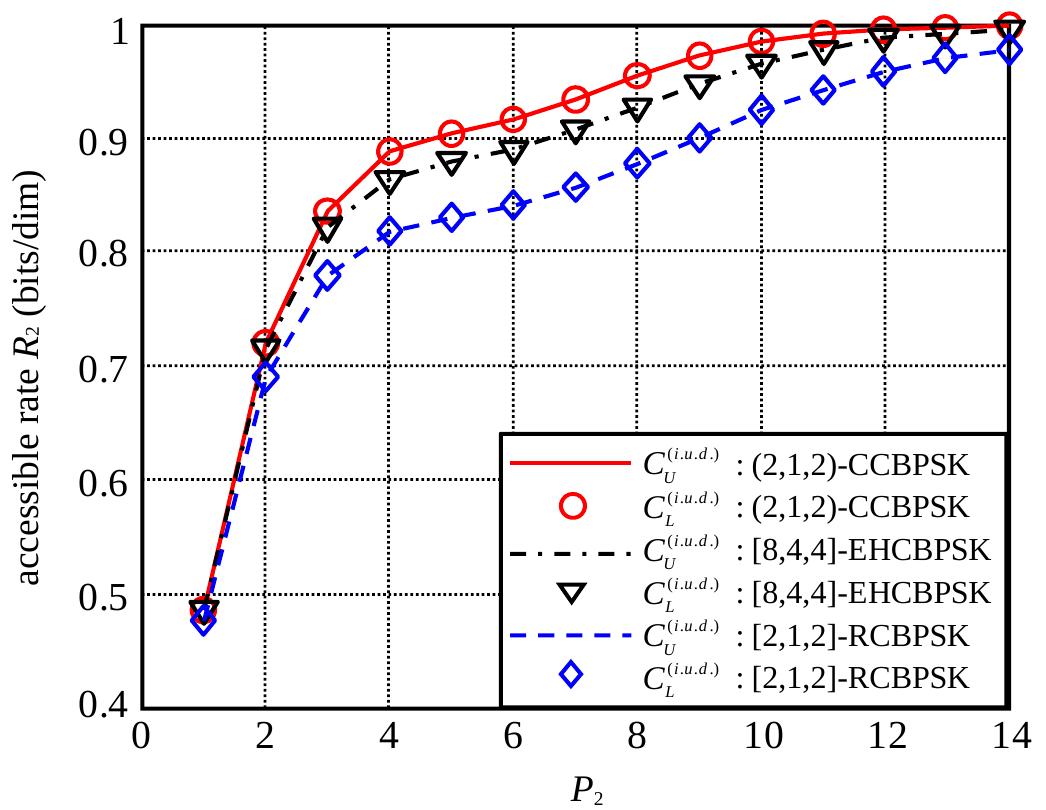}\\
  \caption{Bounds on the i.u.d. accessible capacities of the GIFC with $a_{12}^2 = a_{21}^2 = 1$ for different GTCs with the same coding rate $1/2$.}\label{FigHalfRMid}
\end{figure}

\begin{figure}[!t]
  \centering
  \includegraphics[width=10.0cm]{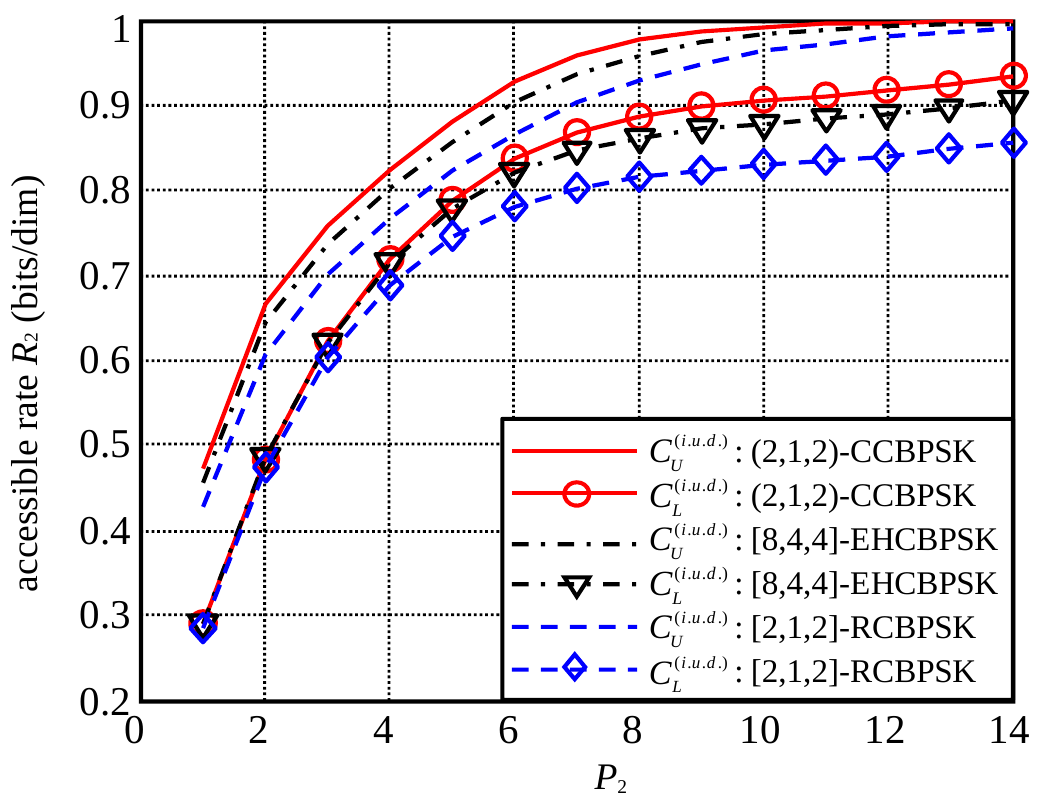}\\
  \caption{Bounds on the i.u.d. accessible capacities of the GIFC with $a_{12}^2 = a_{21}^2 = 0.5$ for different GTCs with the same coding rate $1/2$.}\label{FigHalfRWeak}
\end{figure}

Figs.~\ref{FigUnCoStrong},~\ref{FigUnCoMid} and~\ref{FigUnCoWeak} illustrate the computational results for three different GTCs with {\em different} coding rates over three GIFCs with different interference coefficients, respectively. Figs.~\ref{FigHalfRStrong},~\ref{FigHalfRMid} and~\ref{FigHalfRWeak} illustrate the computational results for three different GTCs of the {\em same} coding rate $1/2$ over three GIFCs with different interference coefficients, respectively.
Recall that the GIFC has been standardized such that the noise power is unit (see~(\ref{eqnIFCseq})). In all plots and tables in this paper, the powers $P_1$ and $P_2$ are measured by the real SNR instead of decibel.
From Figs.~\ref{FigUnCoStrong}-\ref{FigHalfRWeak}, we can see that,
\begin{itemize}
\item {\em primary users with lower coding rates allow higher accessible rates;}
\item {\em given the primary coding rate $R_1$ and the primary transmission power $P_1$, a GTC that provides a better performance allows a higher accessible rate.}
\end{itemize}

%---------------------------------------------------------------------------------
%---------------------------------------------------------------------------------

%---------------------------------------------------------------------------------
%---------------------------------------------------------------------------------

%---------------------------------------------------------------------------------
%---------------------------------------------------------------------------------

%---------------------------------------------------------------------------------
%---------------------------------------------------------------------------------

%---------------------------------------------------------------------------------
%---------------------------------------------------------------------------------

%---------------------------------------------------------------------------------

The above observations can be interpreted intuitively as follows.
Recall that the computed lower bound is derived under the assumptions that i)~the secondary transmitter uses an i.u.d. input, and ii)~the primary receiver employs a successive interference cancellation decoding algorithm.
That is, the secondary message must be decoded correctly with high probability at the primary receiver by treating the ``noise" as a high-dimensional mixed Gaussian noise with ``modes" at ${\bf x}_1 \in \mathscr{C}_1$. In the first case, the lower the primary coding rate~(the fewer codewords used at the primary transmitter), the less crowded the modes are. In the second case, a better GTC~(with lower decoding error probability at the primary receiver in the absence of the secondary users) typically implies that the modes are not that crowded.
In any case, the less crowded primary modes allow the secondary transmitter~(with i.u.d. codewords) to ``insert/superpose'' more distinguishable modes~(secondary codewords), resulting in higher accessible rates.\footnote{It appears that, if the primary modes are squeezed together, ``more space'' would be left for the secondary users. However, given that the mixed-Gaussian noise has~(incompressible) power $1+P_1$, the primary codewords~(if squeezed together) must be distributed non-uniformly over the signal space. If so, the secondary transmitter cannot be benefited from the extra space if the secondary codewords are generated independently and uniformly.}

%Also note that higher decoding error probability at the primary receiver typically implies that the modes are crowded.
%
%The above observations can be interpreted as follows. As pointed out in  [41, 52]£¬ the secondary receiver sees
%a high-dimensional mixed Gaussian noise with ¡°modes¡± at x1 ¡Ê C1.  Intuitively, the more codewords used at the primary transmitter, the more crowded the modes are. Also note that higher decoding error probability at the primary receiver typically implies
%that the modes are crowded.
%
%Recalling that the lower bound is derived under the assumption that the secondary message is decodable at both receivers, we may see why mixed-Gaussian with crowded modes prevents the secondary users from transmitting more messages.
%
%Recalling that the lower bound for no interference margin case is derived under the assumption that the secondary message must be decoded correctly with high probability at the primary receiver, we then see by intuition that mixed-Gaussian with crowded modes prevents the secondary user from transmitting more codewords.

\begin{figure}[!t]
  \centering
  \includegraphics[width=10.0cm]{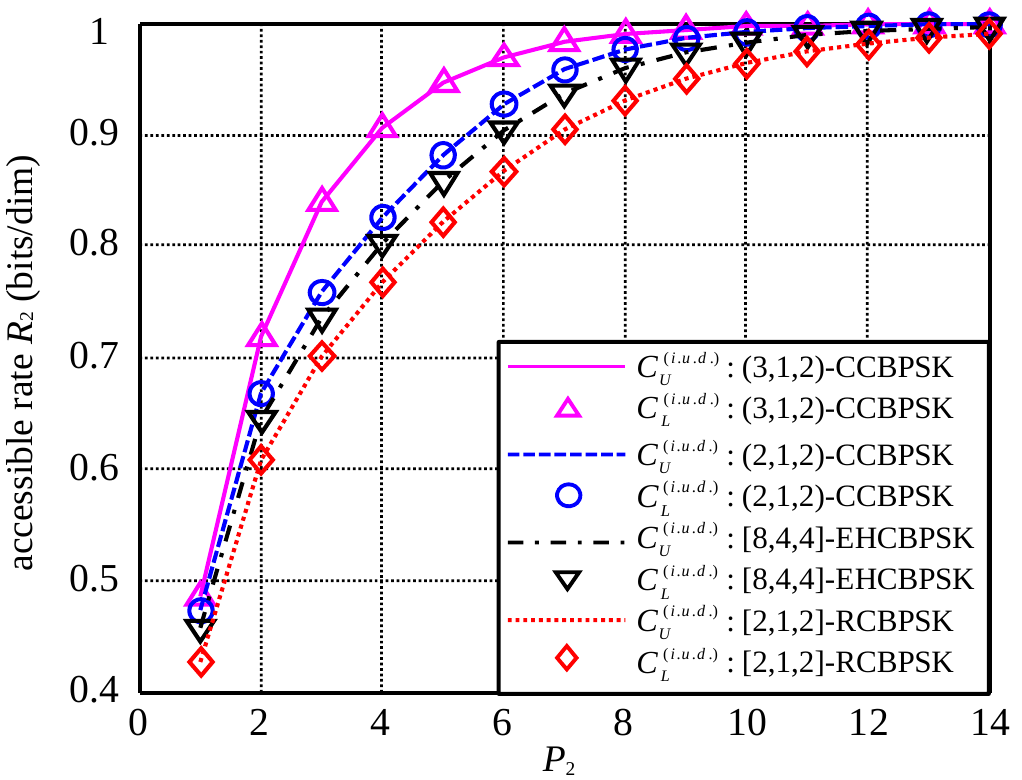}\\
  \caption{Bounds on the i.u.d. accessible capacities of the asymmetric GIFC with $a_{12}^2 = 0.5, a_{21}^2 = 1.5$ for different GTCs.}\label{FigBlockStrong}
\end{figure}

Fig.~\ref{FigBlockStrong} illustrates the computational results for the asymmetric GIFC with $a_{12}^2 = 0.5$ and $a_{21}^2 = 1.5$. From this figure, we can see that the upper bounds coincide with the lower bounds, which verifies Corollary~\ref{CorStrIF} for the case of strong interference at the primary receiver.

%---------------------------------------------------------------------------------
%---------------------------------------------------------------------------------
%---------------------------------------------------------------------------------

%---------------------------------------------------------------------------------
\begin{figure}[!t]
  \centering
  \includegraphics[width=9.0cm]{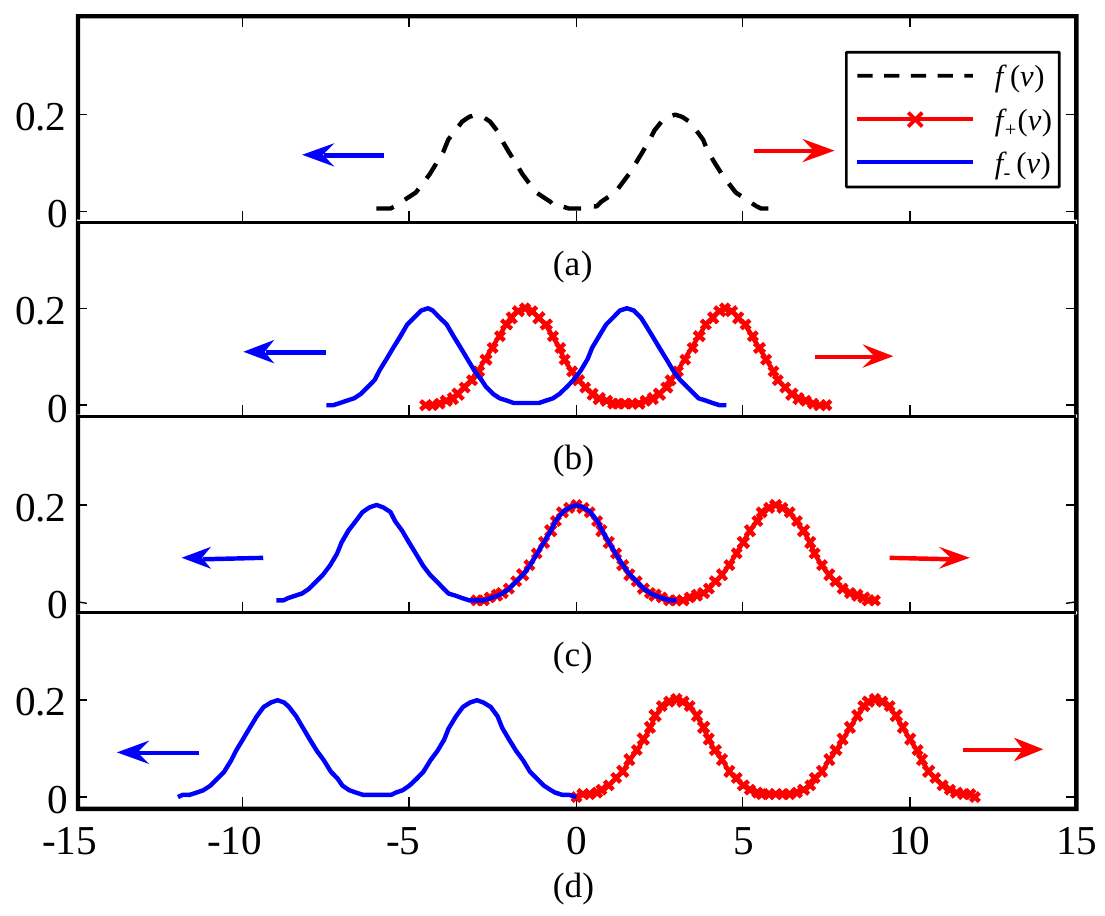}\\
  \caption{The ``noise'' seen by the secondary user~(User D) is the sum of the interference from User A and the Gaussian noise $Z_2$, i.e., $V = a_{12}X_1 + Z_2$. Let $a_{12}^2 = 1.5$ and $X_1$ be drawn from $\{+\sqrt{P_1}, -\sqrt{P_1}\}$ uniformly. In sub-figure~(a), the two-mode curve characterizes the pdf $f(v)$ of $V$. Sub-figure~(b) shows us that, as $P_2$ increases from $0$ to $2$, the two likelihood functions $f_{-}(v) = f(v+\sqrt{P_2})$ and $f_{+}(v) = f(v-\sqrt{P_2})$ become more and more distinguishable. While, sub-figure~(c) shows us that, as $P_2$ continuously increases, the two likelihood functions become less and less distinguishable. Finally, for sufficiently large $P_2$, the two likelihood functions are~(almost) completely distinguishable, as shown in sub-figure~(d).}\label{FigMultNoise}
\end{figure}
%---------------------------------------------------------------------------------
{\bf Remark.} It is worth pointing out that, as seen from the above figures, {\em the accessible rate does not always increase as the secondary transmission power $P_2$ increases}. This is because that, given the primary encoder, the secondary links are no longer AWGN channels. The ``noise" seen by User D, which is composed of interference from the primary user and the Gaussian noise, may have a pdf with multi-modes. Usually, for an additive noise with single-mode pdf, the transmitted signals at the secondary user become more distinguishable as the transmission power increases. However, when the additive noise has a multi-mode pdf, higher transmission power may result in heavier overlaps between the likelihood functions. Let us see an example. Suppose that User A utilizes uncoded BPSK encoder with fixed power $P_1 = 6$. In the case of $a_{12}^2 = 1.5$, the noise seen by User D is $V = a_{12}X_1 + Z_2$, which has a pdf $f(v)$ with two modes as shown in Fig.~\ref{FigMultNoise}~(a). Suppose that User C employs BPSK signaling $\{+\sqrt{P_2}, -\sqrt{P_2}\}$. As $P_2$ increases from $0$ to $2$, we can see that the two likelihood functions $f_{+}(v) = f(v-\sqrt{P_2})$ and $f_{-}(v) = f(v+\sqrt{P_2})$ become more and more distinguishable, as shown in Fig.~\ref{FigMultNoise}~(b). While, as $P_2$ continuously increases, the two likelihood functions become less and less distinguishable, as shown in Fig.~\ref{FigMultNoise}~(c). Finally, as $P_2$ goes to infinity, the two likelihood functions are~(almost) completely distinguishable, as shown in Fig.~\ref{FigMultNoise}~(d).

\subsubsection{The Impact of the Error Performance Requirement on Accessible Rates}

\begin{figure}[!t]
  \centering
  \includegraphics[width=10.0cm]{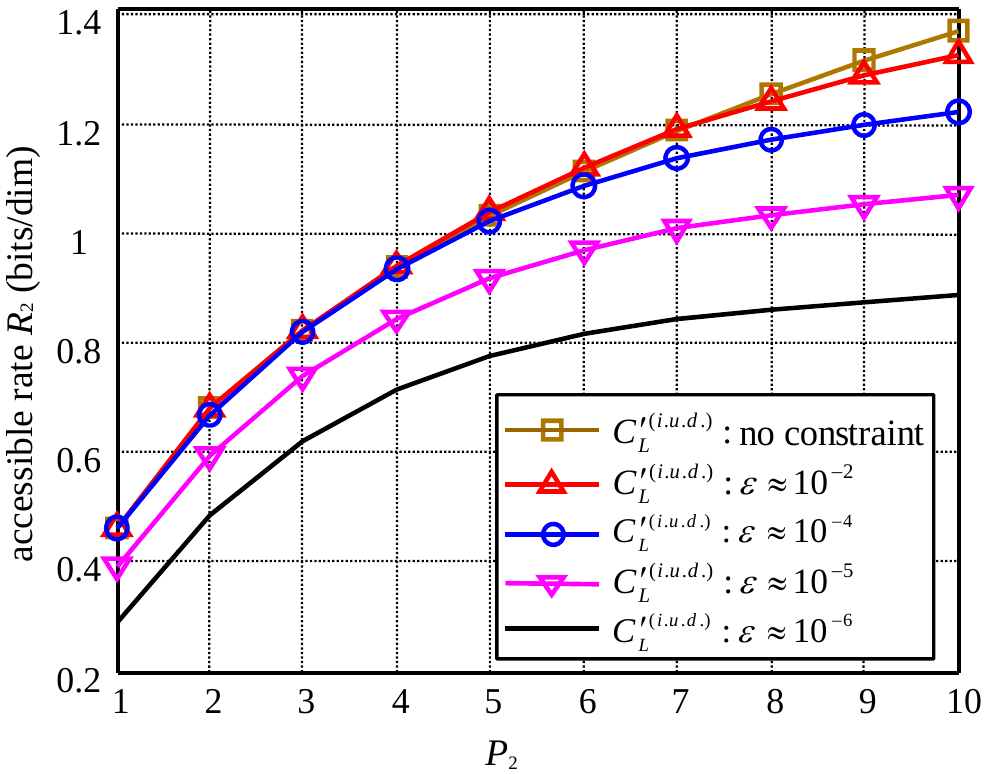}\\
  \caption{Lower bounds on the accessible capacity of the GIFC with $a_{12}^2 = 0.5, a_{21}^2 = 0.5$, when the GTC is fixed as $(2,1,2)$-CCBPSK.}\label{FigLB-epsilon}
\end{figure}

The GTC at User~A is fixed as $(2, 1, 2)$-CCBPSK and the transmission power is fixed as $P_1 = 4.75$.
At the beginning, the error performance requirement by the primary users is set to be $\varepsilon \approx 10^{-5}$ corresponding to the interference margin $\delta_1 = 1.18$, as shown in Table~\ref{TablCompEPR}. This error performance requirement can be either strengthened by decreasing $\varepsilon\approx 10^{-5}$ down to $\varepsilon_{\min}\approx 10^{-6}$ corresponding to $\delta_1=1$ (shown in Table~\ref{TablCompEPR}) or relaxed by increasing $\varepsilon$ from $10^{-5}$ to $10^{-4}$ and $10^{-2}$ resulting in larger interference margins (shown in Table~\ref{TablCompEPR}). In all these cases, the lower bounds ${C'}_L^{(i.u.d.)}$ given in~(\ref{eqnLBnon-trivIUD}) can be computed, as shown in Fig.~\ref{FigLB-epsilon}. Also shown in Fig.~\ref{FigLB-epsilon} is the curve of the lower bounds ${C'}_L^{(i.u.d.)}$ when the error performance requirement by the primary users is not considered at all, which is referred to as ``no-constraint''. From this figure, we can see that {\em relaxing the error performance requirement at the primary users allows the secondary users to have higher accessible rates}. However, the rates must be upper bounded by the case where no error performance requirement by the primary users is constrained.
%We can also see from this figure that {\em these computable lower bounds are bounded from both below and above}.
%
%The error performance requirement by the primary users is set to be $\varepsilon \approx 10^{-5}$. One one hand, the error performance requirement can be decreased down to $\varepsilon_{\min} \approx 10^{-6}$ corresponding to $\delta_1=1$. On the other hand, the error performance requirement is relaxed from $10^{-5}$ to $10^{-4}$ and $10^{-2}$ resulting in larger interference margins. In all these cases, the lower bounds ${C'}_L^{(i.u.d.)}$ given in~(\ref{eqnLBnon-trivIUD}) can be computed. Here we simulate only the GIFC with weak interference $a_{12}^2 = a_{21}^2 = 0.5$. Different error performance requirements and other parameters are listed in Table~\ref{TablCompEPR}.

\begin{table}[!t]
  \centering
  \caption{Parameters for the GTC $(2, 1, 2)$-CCBPSK}\label{TablCompEPR}
  \begin{tabular}{|c|l|c|c|}
    \hline
    $\varepsilon$ & $P_1^*$ & $P_1$ & $\delta_1$  \\ \hline
    $\approx 10^{-2}$ & 1.7 & 4.75 & 2.79 \\
    $\approx 10^{-4}$ & 3.2 & 4.75 & 1.48 \\
    $\approx 10^{-5}$ & 4 & 4.75 & 1.18 \\
    $\approx 10^{-6}$ & 4.75 & 4.75 & 1 \\
    \hline
  \end{tabular}
\end{table}

%\begin{table}
%  \centering
%  \caption{Parameters for the GTC $(2, 1, 2)$-CCBPSK}\label{TablCompEPR}
%  \begin{tabular}{|c|c|c|c|}
%    \hline
%    % after \\: \hline or \cline{col1-col2} \cline{col3-col4} ...
%    $\varepsilon$ & $P_1^*$ & $P_1$ & $\delta_1$  \\ \hline
%    $\approx 10^{-2}$ & 1.70 & 4.75 & 2.7941 \\
%    $\approx 10^{-4}$ & 3.20 & 4.75 & 1.4844 \\
%    $\approx 10^{-6}$ & 4.75 & 4.75 & 1 \\
%    \hline
%  \end{tabular}
%\end{table}
%
%Fig.~\ref{FigLB-epsilon} depicts the computational results of lower bounds ${C'}_L^{(i.u.d.)}$ (see~(\ref{eqnLBnon-trivIUD})) on the accessible capacity. From this figure, we can see that {\em relaxing the error performance requirement at the primary users allows the secondary users to have higher accessible rates.}

%---------------------------------------------------------------------------------
%---------------------------------------------------------------------------------

\section{Conclusion}\label{SecConclusion}

%In this paper, we have presented a new problem formulation for the GIFC with primary users and secondary users. We
%defined the accessible capacity as the maximum rate at which the secondary users can communicate reliably without affecting the error performance requirement by the primary users. By modeling the primary encoder as a generalized trellis code~(GTC), we have derived upper and lower bounds on the accessible capacity. We have also revealed the relationships between the accessible capacity and the conventional capacity region of the GIFC.  For some special cases, these bounds can be computed by the BCJR algorithm. The numerical results show us that, either as expected or interestingly,
%\begin{itemize}
%    \item at a fixed coding rate $R_1$, better primary encoders guarantee not only higher quality of the primary links but also higher accessible rates of the secondary users;
%    \item primary users with lower transmission rates may allow higher accessible rates;
%    \item the accessible capacity does not always increase with the transmission power of the secondary transmitter.
%\end{itemize}

In this paper, we have presented a new problem formulation for the GIFC with primary users and secondary users.
We defined the accessible capacity as the maximum rate at which the secondary users can communicate reliably without affecting the error performance requirement by the primary users. Upper and lower bounds on the accessible capacity were derived and evaluated using the BCJR algorithm. Numerical results were also provided to illustrate the dependence of the accessible capacity on parameters.

\appendix[The Discrete Finite State Channel and Theorem~4.6.1 in~\cite{Gallager68}]\label{AppendixFSCcodingTH}

%\section{The Discrete Finite State Channel and Theorem~4.6.1 in~\cite{Gallager68}}\label{AppendixFSCcodingTH}

In this appendix, we re-state Gallager's Theorem~4.6.1~\cite{Gallager68}, which is used to prove Lemma~2 in Sec.~\ref{SubsecTriIM}.

In~\cite{Gallager68}, a discrete finite state channel has an input sequence ${\bf x} = \cdots x_{-1}, x_0, x_1, \cdots$, an output sequence ${\bf y} = \cdots y_{-1}, y_0, y_1, \cdots$, and a state sequence ${\bf s} = \cdots s_{-1}, s_0, s_1, \cdots$. Each input letter $x_n$, each output letter $y_n$ and each state letter $s_n$ are selected from finite alphabets $\{0,1,\ldots,K-1\}$, $\{0,1,\ldots,J-1\}$ and $\{0,1,\ldots,A-1\}$, respectively. The channel is described statistically by the time-invariant conditional probability assignment $p(y_n,s_n|x_n,s_{n-1})$ satisfying
\begin{equation}
    p(y_n,s_n|x^n,y^{n-1},s^{n-1}) = p(y_n,s_n|x_n,s_{n-1}).
\end{equation}
The probability of a given output sequence ${\bf y}=(y_1, \cdots, y_N)$ and a final state $s_N$ at time $N$ conditional on an input sequence ${\bf x}=(x_1, \cdots, x_N)$ and an initial state $s_0$ at time $0$ can be calculated inductively from
\begin{equation}
p_N({\bf y},s_N|{\bf x},s_0) = \sum_{s_{N-1}} p(y_N,s_N|x_N,s_{N-1}) p_{N-1}({\bf y}_{N-1},s_{N-1}|{\bf x}_{N-1},s_0)
\end{equation}
where ${\bf x}_{N-1} = (x_1, \ldots, x_{N-1})$ and ${\bf y}_{N-1} = (y_1, \ldots, y_{N-1})$. The final state can be summed over to give
\begin{equation}
p_N({\bf y}|{\bf x},s_0) = \sum_{s_{N}} p_N({\bf y},s_N|{\bf x},s_0).
\end{equation}

Define
\begin{equation}
\underline{C}_N = \frac{1}{N} \max_{p_N({\bf x})}\min_{s_0} I\left({\bf X};{\bf Y}|s_0\right)
\end{equation}
\begin{equation}
\overline{C}_N = \frac{1}{N} \max_{p_N({\bf x})}\max_{s_0} I\left({\bf X};{\bf Y}|s_0\right).
\end{equation}

{\em Theorem~4.6.1 in~\cite{Gallager68}:} For the above finite state channel with $A$ states,
\begin{equation}
\lim_{N\rightarrow \infty} \underline{C}_N = \sup_{N}\left[\underline{C}_N - \frac{\log A}{N}\right]
\end{equation}
\begin{equation}
\lim_{N\rightarrow \infty} \overline{C}_N = \inf_{N}\left[\overline{C}_N + \frac{\log A}{N}\right].
\end{equation}
\hfill \ding{113}

\section*{Acknowledgment}

The authors would like to thank Prof. David Tse and Prof. Raymond Yeung for their helpful suggestions. The authors are also grateful to reviewers for their valuable comments. The authors also wish to thank Chulong Liang for his help.

\ifCLASSOPTIONcaptionsoff
  \newpage
\fi



\bibliographystyle{IEEEtran}
\bibliography{IEEEabrv,mybibfile}

%\begin{biography}[{\includegraphics[width=1in,height=1.25in,clip,keepaspectratio]{mshell}}]{Michael Shell}

%\end{IEEEbiography}


%\newpage



%\vfill


\end{document}